\documentclass[aps,prd,onecolumn,groupedaddress,nofootinbib,superscriptaddress]{revtex4}  
\usepackage{graphicx}
\usepackage{epstopdf}
\usepackage{amsmath}
\usepackage{amsfonts}
\usepackage{amssymb}
\usepackage{appendix}
\usepackage{comment}
\usepackage{bbold}
\usepackage{color}
\usepackage{slashed}
\usepackage{subfigure}
\usepackage{setspace}
\usepackage{footnote}
\usepackage{multirow}
\usepackage{braket}
\usepackage[normalem]{ulem}
\usepackage[a4paper, portrait, margin=0.5in]{geometry}
\usepackage[utf8]{inputenc}
\usepackage{mathrsfs}
\usepackage{longtable}

\LTcapwidth=\textwidth

\usepackage{url}
\usepackage{wrapfig}
\usepackage{braket}

\newcommand{\GFU}{\left(\frac{\sqrt{2}}{G_F}\right)}
\newcommand{\GF}{\left(\frac{G_F}{\sqrt{2}}\right)}

\newcommand{\qPU}{\left(q^2\right)}
\newcommand{\qP}{\left(\frac{1}{q^2}\right)}

\newcommand{\TBB}{T_{0\nu\beta\beta}}
\newcommand{\RME}{R_{\mu^- e^+}}

\begin{document}
\singlespacing
{\hfill NUHEP-TH/16-07}

\title{On Lepton-Number-Violating Searches for Muon to Positron Conversion}

\author{Jeffrey M. Berryman}
\affiliation{Northwestern University, Department of Physics \& Astronomy, 2145 Sheridan Road, Evanston, IL 60208, USA}
\author{Andr\'{e} de Gouv\^{e}a}
\affiliation{Northwestern University, Department of Physics \& Astronomy, 2145 Sheridan Road, Evanston, IL 60208, USA}
\author{Kevin J. Kelly}
\affiliation{Northwestern University, Department of Physics \& Astronomy, 2145 Sheridan Road, Evanston, IL 60208, USA}
\author{Andrew Kobach}
\affiliation{Department of Physics, University of California, San Diego, La Jolla, CA 92093, USA}

\begin{abstract}
There is no guarantee that the violation of lepton number, assuming it exists, will primarily manifest itself in neutrinoless double beta decay ($0\nu \beta \beta$). Lepton-number violation and lepton-flavor violation may be related, and much-needed information regarding the physics that violates lepton number can be learned by exploring observables that violate lepton flavors other than electron flavor. One of the most promising observables is $\mu^-\rightarrow e^+$ conversion, which can be explored by experiments that are specifically designed to search for $\mu^- \rightarrow e^-$ conversion.  We survey lepton-number violating dimension-five, -seven, and -nine effective operators in the standard model and discuss the relationships between Majorana neutrino masses and the rates for $0\nu \beta\beta$ and $\mu^- \rightarrow e^+$ conversion. While  $0\nu \beta\beta$ has the greatest sensitivity to new ultraviolet energy scales, its rate might be suppressed by the new physics relationship to lepton flavor, and $\mu^-\rightarrow e^+$ conversion offers a complementary probe of lepton-number-violating physics.
\end{abstract}

\maketitle

\section{Introduction}
\label{sec:Introduction}

Neutrino flavor oscillations imply that at least two neutrinos have nonzero masses and that there is nontrivial mixing in the lepton sector. The mechanism behind nonzero neutrino masses is currently unknown, and a definitive resolution of the neutrino mass puzzle will require input from a variety of probes of fundamental physics, including neutrino oscillation experiments, searches for lepton-number and baryon-number violation, precision measurements of charged-lepton properties and rare processes, and high-energy collider experiments.

Tests of the validity of lepton-number conservation are among the most valuable sources of information when it comes to the neutrino mass puzzle (see, for example, Ref.~\cite{deGouvea:2013zba}, for an overview). They provide unique information on the nature of the neutrino, i.e., whether it is a Dirac or Majorana fermion. Speculations on the origin of neutrino masses, in turn, differ dramatically depending on the nature of the neutrino. While searches for neutrinoless double beta decay ($0\nu\beta\beta$) are, by far, the most powerful available probes of lepton-number violation (see Ref.~\cite{Rodejohann:2011mu} for a thorough overview), the pursuit of other lepton-number-violating (LNV) observables is of the highest importance. 

Searches for charged-lepton-flavor violation are also potentially powerful probes of the origin of neutrino masses (see, for example, Ref.~\cite{deGouvea:2013zba,Lindner:2016bgg}, for an overview). Among the different charged-lepton-flavor-violating processes, powerful new searches for to $\mu^-\to e^-$ conversion in nuclei are currently being developed \cite{Cui:2009zz,Natori:2014yba,Bartoszek:2014mya}. These are expected to improve on the current sensitivity to the $\mu^-\to e^-$ conversion rate by at least four orders of magnitude in less than a decade. 

Experiments sensitive to $\mu^-\to e^-$ conversion in nuclei may also serve as laboratories to search for the LNV $\mu^-\to e^+$ conversion in nuclei (see, for example, Ref.~\cite{Bartoszek:2014mya,Kuno:2015tya}). The current upper bound on this conversion rate, normalized to the capture rate, is $1.7\times 10^{-12}$ for the transition between titanium and the ground state of calcium, obtained by the SINDRUM II collaboration \cite{Kaulard:1998rb}. Significant improvement is expected from at least a subset of the next-generation $\mu^-\to e^-$ conversion experiments. 

Here, we estimate the capabilities of next-generation $\mu^-\to e^-$ conversion experiments to discover or constrain $\mu^-\to e^+$ conversion in nuclei. We also explore how these results can relate to searches for $0\nu\beta\beta$ and nonzero Majorana neutrino masses.  We make use of the standard model (SM) effective operator approach -- introduced in Ref.~\cite{Babu:2001ex} and explored in, for example, Refs.~\cite{deGouvea:2007qla,Angel:2012ug} -- in order to gauge the impact of these future measurements on a large variety of neutrino mass models. This approach is powerful, and allows one to relate different LNV observables, including nonzero neutrino masses. Extended versions of this approach have been successfully pursued in order to relate, assuming grand unification is realized in nature, lepton-number and baryon-number violating observables \cite{deGouvea:2014lva}. For other comparisons of $\mu^-\to e^+$ conversion in nuclei to different LNV observables see, for example, Refs.~\cite{Atre:2005eb,Rodejohann:2011mu,Geib:2016atx}. Ref.~\cite{Geib:2016atx}, which appeared in the literature shortly before this work, asks some of the questions we address here, but our approaches are somewhat complementary. More concretely, we analyze LNV phenomena using a different set of effective operators, as will be explained below.

This manuscript is organized as follows. In Sec.~\ref{sec:exp}, we estimate the sensitivity of different next-generation $\mu^-\to e^-$ conversion experiments to $\mu^-\to e^+$ conversion in nuclei. In Sec.~\ref{sec:theory}, we review the effective operator approach and identify the operators of interest. We also review how the mass scale of the different effective operators can be related to the observed neutrino masses. In Sec.~\ref{sec:Estimates}, we discuss a few concrete examples of how we estimate the rates for the LNV processes of interest, and in Sec.~\ref{sec:Results}, we present and discuss our results. We present some concluding thoughts in Sec.~\ref{sec:Conclusions}.


\section{Sensitivities of Next-Generation Experiments}
\label{sec:exp}

The SINDRUM II experiment at PSI was designed to investigate $\mu^- \to e^-$ conversion in nuclei. The most recent result places a limit on $\mu^- \to e^-$ conversion in gold \cite{Bertl:2006up},
\begin{equation}
 \label{S2minus}
R_{\mu^- e^-}^\text{Au} \equiv \frac{\Gamma(\mu^- + \text{Au} \to e^- + \text{Au})}{\Gamma(\mu^- + \text{Au} \to \nu_\mu + \text{Pt})} < 7 \times 10^{-13} \text{ (90\% CL).}
\end{equation}
Nearly ten years earlier, the SINDRUM II collaboration also set a limit on $\mu^- \to e^+$ conversion in titanium \cite{Kaulard:1998rb}, 
\begin{equation}
\label{S2plus}
R_{\mu^- e^+}^\text{Ti} \equiv \frac{\Gamma(\mu^- + \text{Ti} \to e^+ + \text{Ca})}{\Gamma(\mu^- + \text{Ti} \to \nu_\mu + \text{Sc})} < \left\{
\begin{array}{l}
1.7 \times 10^{-12} \text{ (GS, 90\% CL)} \\ 
3.6 \times 10^{-11} \text{ (GDR, 90\% CL)}  \end{array} \right. ,
\end{equation}
where the top limit (GS) assumes coherent scattering to the ground state of calcium, while the bottom limit (GDR) assumes a transition to a giant dipole resonance state. The GS limit remains the strongest on any $\mu^- \to e^+$ conversion process to date. Next-generation experiments, however, are expected to improve upon it by several orders of magnitude.

The next generation of $\mu^- \to e^-$ conversion experiments includes Mu2e \cite{Bartoszek:2014mya} at Fermilab in the U.S. and DeeMe \cite{Natori:2014yba} and COMET \cite{Cui:2009zz} (and its upgrade, PRISM \cite{Kuno:2012pt}) at J-PARC in Japan. Mu2e and COMET/PRISM are schematically similar to SINDRUM II: a proton beam impinges upon a pion production target, and the muons produced in the pion decays are directed onto an aluminum stopping target. DeeMe is similar to these, except the pion production, muon production and muon capture all take place in the same SiC target. The muons form bound states with the atomic nuclei, at which point one of the following happens: (i) the muons decay in orbit (DIO); (ii) they are captured by the nucleus, and a neutrino is produced; or (iii) they interact with the nucleus in a way not prescribed by the SM. DIO is one of the largest backgrounds at these experiments; the endpoint of the DIO electron spectrum coincides with the energy of the electron produced in $\mu^- \to e^-$ conversion. The spectrum of DIO electrons is calculable, however, and any unaccounted-for electrons in the region $E_e \sim m_\mu$ would constitute a signal. These experiments anticipate the following sensitivities ($R^{\text{SiC}}_{\mu^- e^-}$ and $R^{\text{Al}}_{\mu^- e^-}$ are defined analogously to Eq.~\eqref{S2minus}):
\begin{align*}
\text{DeeMe:} & \quad R^{\text{SiC}}_{\mu^- e^-} > 5 \times 10^{-14} \text{ (90\% CL)}, \\
\text{Mu2e:} & \quad R^{\text{Al}}_{\mu^-e^-} >  6.6 \times 10^{-17} \text{ (90\% CL)}, \\
\text{COMET Phase-I:} & \quad R^{\text{Al}}_{\mu^-e^-} > 7.2 \times 10^{-15} \text{ (90\% CL)}, \\
\text{COMET Phase-II:} & \quad R^{\text{Al}}_{\mu^-e^-} > 6 \times 10^{-17} \text{ (90\% CL)}, \\
\text{PRISM:} & \quad R^{\text{Al}}_{\mu^-e^-} > 5 \times 10^{-19} \text{ (90\% CL)}.
\end{align*}

Here we qualitatively estimate the sensitivities of these experiments to $\mu^- \to e^+$ conversion. At DeeMe, COMET Phase-II, and PRISM, the electrons ejected from the stopping target are transported away from the target to the spectrometer via magnetic fields. This helps to reject background events, but also means that, naively, any produced positrons will be ``swept away'' and not detected, rendering  $\mu^- \to e^+$ conversion searches significantly more challenging and potentially unfeasible. We are therefore not able to infer a sensitivity for these experiments. Mu2e and COMET Phase-I, however, are a different story. The aluminum stopping targets are immersed in an external magnetic field, and the energies of emitted electrons are measured by determining their trajectories after they escape the stopping target. These experiments can then directly determine if an emitted lepton is an electron or a positron. This is precisely how limits on $R_{\mu^-e^+}$ were determined at SINDRUM II. We estimate the sensitivities of these experiments to $\mu^- \to e^+$ conversion as follows. In Ref.~\cite{Dohmen:1993mp}, the SINDRUM II collaboration set limits on $R^{\text{Ti}}_{\mu^- e^-}$ and $R^{\text{Ti}}_{\mu^- e^+}$ (assuming transitions to the ground state of calcium) for the same experimental run:
\begin{align*}
\label{S2limits}
R^{\text{Ti}}_{\mu^- e^-} < 4.3 \times 10^{-12} \text{ (90\% CL)}, \\
 R^{\text{Ti}}_{\mu^- e^+} < 4.3 \times 10^{-12} \text{ (90\% CL)}.
\end{align*}
(That these two bounds are identical is a numerical accident.) Since these two limits are quite comparable to each other, we assume the improvements in the sensitivities to $\mu^- \to e^-$ conversion and $\mu^- \to e^+$ conversion scale commensurately and estimate that next-generation experiments will be sensitive to $\mu^- \to e^+$ rates greater than the following:
\begin{align*}
\text{Mu2e:} & \quad R^{\text{Al}}_{\mu^-e^+} \gtrsim 10^{-16}, \\
\text{COMET Phase-I:} & \quad R^{\text{Al}}_{\mu^-e^+} \gtrsim 10^{-14}.
\end{align*}
We emphasize that these are crude estimates. Detailed experimental analyses of the sensitivities of these experiments to $\mu^- \to e^+$ conversion do not exist in the literature and a realistic estimate can only be made in association with the existing experimental collaboration. We echo the sentiment recently expressed by the authors of Ref.~\cite{Geib:2016atx}, that such analyses should be pursued as they can potentially play a significant role in the study of LNV phenomena.


\section{Effective Operator Approach}
\label{sec:theory}

The SM Lagrangian can be augmented by operators with mass-dimension $d>4$, that are constructed from SM matter fields ($Q$, $u^c$, $d^c$, $L$, $e^c$), Higgs bosons ($H$), field strength tensors ($G_{\mu\nu}$, $W_{\mu\nu}$, $B_{\mu\nu}$) and covariant derivatives ($D_\mu$) (and their complex conjugates) and that respect both gauge and Lorentz invariance. These operators, however, need not respect the global symmetries of baryon number and lepton number. LNV phenomena, including $0\nu\beta\beta$ and neutrino Majorana masses, arise from operators that violate lepton number by two units $(\Delta L = \pm 2)$ and conserve baryon number $(\Delta B = 0)$. It was recently proven in Ref.~\cite{Kobach:2016ami}, and considered earlier in Refs.~\cite{Rao:1983sd,deGouvea:2014lva}, that operators in the SM with $|\Delta B-\Delta L|=2$ must have odd mass-dimension. The operators included in our analysis are listed in Tables~\ref{EstimateTableD5} (dimension-five), \ref{EstimateTableD7} (dimension-seven), and \ref{EstimateTableD9} (dimension-nine). We consider operators with $d \le 9$ that contain neither (covariant) derivatives nor field strength tensors; the number of operators with $|\Delta L| = 2$ grows quickly when $d \ge 11$ (see Refs.~\cite{Babu:2001ex,deGouvea:2007qla,Angel:2012ug}). Fields whose $SU(2)_L$ indices are contracted to form singlets are enclosed in parentheses; operators with the same field content but with different $SU(2)_L$ structure are listed separately. An operator may have multiple possible contractions of its $SU(3)_c$ and Lorentz indices. However, these different contractions lead to very similar estimates  -- they differ by at most $\mathcal{O}(1)$ -- for the amplitudes of LNV processes of interest here and will henceforth be ignored. 

As already mentioned in the introduction, the effective operator approach employed here is complimentary to the analyses in Ref.~\cite{Geib:2016atx}, in which a different set of effective operators is used. Specifically, the operators of Ref.~\cite{Geib:2016atx} are constructed to be invariant under the low-energy symmetry group $SU(3)_c \times U(1)_{EM}$ as opposed to the full SM gauge group. Fig. 4 of that paper depicts experimental limits and sensitivities for the Wilson coefficient of the operator
\begin{equation}
\label{merle-op}
\left( \overline{d} \gamma^\mu P_L u \right) \left( \overline{d} \gamma_\mu P_L u \right) \left( \overline{e^c} P_L \ell \right), \quad \ell = e, \, \mu. 
\end{equation}
This low-energy operator is descended from the following SM-gauge-invariant operators:
\begin{eqnarray}
\label{dim11ops}
\mathcal{O}_{47_a} & = & (L\overline{Q}) (L\overline{Q}) (H Q) (H Q), \\
\mathcal{O}_{47_d} & = & (L\overline{Q}) (LQ) (H Q) (H\overline{Q}), 
\end{eqnarray}
where we have used the naming convention of Ref.~\cite{deGouvea:2007qla,Angel:2012ug}. These operators have mass-dimension eleven, and thus lie outside the scope of this work.

\begin{table}[t]
\caption{The dimension-five operator featured in this analysis. Naming convention follows from Refs.~\cite{deGouvea:2007qla,Angel:2012ug}. Parentheses denote fields that have their $SU(2)_L$ indices contracted to form a singlet. While not explicitly indicated, three generations of all fermions are contained in each operator. In the third column, $\Lambda$ is the scale required to produce a neutrino mass in the range $0.05 - 0.5$ eV, with lower $\Lambda$ corresponding to higher neutrino mass. Analytic estimates of $\TBB$ and $\RME$ are also listed, along with numerical estimates, assuming the operator in question is responsible for the observable neutrino masses. See text for details.}
\begin{center}
\begin{tabular}{ |c|c|c||l| }
\hline
\multirow{2}{*}{$\mathcal{O}$} & \multirow{2}{*}{Operator} & \multirow{2}{*}{$\Lambda$ [TeV]} & $T_{0\nu\beta\beta}$ \\[5pt] \cline{4-4}
& & & $R_{\mu^- e^+}$ \\[5pt] \hline \hline

\multirow{2}{*}{$\mathcal{O}_1$} & \multirow{2}{*}{$(LH) (LH)$} & \multirow{2}{*}{$6\times 10^{10-11}$} & $\ln{(2)} \GFU^4 \qPU^2 \frac{1}{v^4} \frac{\Lambda^2}{Q^{11}} \sim 10^{25}-10^{27}$ yr \\[5pt] \cline{4-4}
& & &  $\GF^2 \qP^2 \frac{v^4 Q^6}{\Lambda^2} \sim 10^{-38} - 10^{-36}$ \\[5pt] \hline
\end{tabular}
\end{center}
\label{EstimateTableD5}
\end{table}

\begin{table}[t]
\caption{Same as Table~\ref{EstimateTableD5}, for the dimension-seven operators featured in this analysis. Naming convention follows from Refs.~\cite{deGouvea:2007qla,Angel:2012ug}.}
\begin{center}
\begin{tabular}{ |c|c|c||l| }
\hline
\multirow{2}{*}{$\mathcal{O}$} & \multirow{2}{*}{Operator} & \multirow{2}{*}{$\Lambda$ [TeV]} & $T_{0\nu\beta\beta}$ \\[5pt] \cline{4-4}
& & & $R_{\mu^- e^+}$ \\[5pt] \hline \hline

\multirow{2}{*}{$\mathcal{O}_2$} & \multirow{2}{*}{$(LL) (LH) e^c$} & \multirow{2}{*}{$4\times 10^{6-7}$} & $\ln{(2)} \GFU^4 \qPU^2 \left(\frac{16\pi^2}{y_\tau v^2}\right)^2 \frac{\Lambda^2}{Q^{11}} \sim 10^{25}-10^{27}$ yr \\[5pt] \cline{4-4}
& & & $ \GF^2 \qP^2 \left(\frac{y_\tau v^2}{16\pi^2}\right)^2 \frac{Q^6}{\Lambda^2} \sim 10^{-38}-10^{-36}$ \\[5pt] \hline

\multirow{2}{*}{$\mathcal{O}_{3_a}$} & \multirow{2}{*}{$(LL) (QH) d^c$} & \multirow{2}{*}{$2\times 10^{4-5}$} & $\ln{(2)} \GFU^2 q^2 \frac{\Lambda^2}{Q^{11}} \left[ \GF^2 \frac{1}{q^2} \left(\frac{ y_b g^2 v^2}{(16\pi^2)^2}\right)^2 + \frac{v^2}{\Lambda^4} \right]^{-1} \sim 10^{24}-10^{26}$ yr \\[5pt] \cline{4-4}
& & & $\frac{1}{q^2}\frac{Q^6}{\Lambda^6} \left[ \GF^2 \frac{1}{q^2} \left(\frac{y_b g^2 v^2}{(16\pi^2)^2}\right)^2 + \frac{v^2}{\Lambda^4}\right] \sim 10^{-37}-10^{-36}$ \\[5pt] \hline

\multirow{2}{*}{$\mathcal{O}_{3_b}$} & \multirow{2}{*}{$(LQ)(LH)d^c$} & \multirow{2}{*}{$1 \times 10^{7-8}$} & $\ln{(2)}\GFU^2 q^2 \frac{\Lambda^2}{Q^{11}} \left[ \GF^2 \frac{1}{q^2} \left(\frac{y_b v^2}{16\pi^2}\right)^2 + \frac{v^2}{\Lambda^4}\right]^{-1} \sim 10^{25}-10^{27}$ yr \\[5pt] \cline{4-4}
& & & $\frac{1}{q^2} \frac{Q^6}{\Lambda^2} \left[\GF^2 \frac{1}{q^2}\left(\frac{y_b v^2}{16\pi^2}\right)^2 + \frac{v^2}{\Lambda^4} \right] \sim 10^{-38}-10^{-36}$\\[5pt] \hline

\multirow{2}{*}{$\mathcal{O}_{4_a}$} & \multirow{2}{*}{$(L\overline{Q})(LH) \overline{u^c}$} & \multirow{2}{*}{$4\times 10^{8-9}$} & $\ln{(2)} \GFU^2 q^2 \frac{\Lambda^2}{Q^{11}} \left[ \GF^2 \frac{1}{q^2} \left(\frac{y_t v^2}{16\pi^2}\right)^2 + \frac{v^2}{\Lambda^4} \right]^{-1} \sim 10^{25}-10^{27}$ yr \\[5pt] \cline{4-4}
& & & $\frac{1}{q^2}\frac{Q^6}{\Lambda^2} \left[ \GF^2 \frac{1}{q^2} \left(\frac{y_t v^2}{16\pi^2}\right)^2 + \frac{v^2}{\Lambda^4}\right] \sim 10^{-38}-10^{-36}$ \\[5pt] \hline

\multirow{2}{*}{$\mathcal{O}_{4_b}$} & \multirow{2}{*}{$(LL)(\overline{Q}H)\overline{u^c}$} & \multirow{2}{*}{$2-7$} & This operator can not contribute to $0\nu\beta\beta$. \\[5pt] \cline{4-4}
& & & $\frac{v^2 Q^6}{q^2 \Lambda^2} \left[\GF^2 \left(\frac{y_d}{g^2}\right)^2 \left(\frac{y_t}{16\pi^2}\right)^2 + \frac{1}{\Lambda^4}\right] \sim 10^{-27}-10^{-24}$ \\[5pt] \hline

\multirow{2}{*}{$\mathcal{O}_{8}$} & \multirow{2}{*}{$(LH)\overline{e^c}\overline{u^c}d^c$} & \multirow{2}{*}{$6\times 10^{2-3}$} & $\ln{(2)}\GFU^2 q^2 \frac{\Lambda^2}{Q^{11}}\left[ \GF^2 \frac{1}{q^2}\left(\frac{y_t y_b v}{(16\pi^2)^2}\right)^2 + \frac{v^2}{\Lambda^4}\right]^{-1} \sim 10^{27}-10^{29}$ yr \\[5pt] \cline{4-4}
& & & $\frac{1}{q^2}\frac{Q^6}{\Lambda^2}\left[ \GF^2 \frac{1}{q^2} \left(\frac{y_t y_b y_\mu v^2}{(16\pi^2)^2}\right)^2 + \frac{v^2}{\Lambda^4}\right] \sim 10^{-40}-10^{-38}$ \\[5pt] \hline
\end{tabular}
\end{center}
\label{EstimateTableD7}
\end{table}

In the absence of neutrino masses, the SM exhibits global $U(1)$ symmetries associated with each lepton flavor, electron-number, muon-number, and tau-number. This is no longer the case in the presence of beyond-the-standard-model physics, and lepton-flavor numbers are necessarily violated if global lepton number is violated. The operators in Tables~\ref{EstimateTableD5}, \ref{EstimateTableD7}, and \ref{EstimateTableD9} can distribute their lepton-number violation between the lepton families. For instance, the Weinberg operator $\mathcal{O}_1$ should be generalized to
\begin{equation}
\label{definef}
\frac{1}{\Lambda} (L H) (L H) \to \frac{1}{\Lambda} \Big[ f_{ee} (L_e H) (L_e H) + f_{e\mu} (L_e H) (L_\mu H) + f_{e\tau} (L_e H) (L_\tau H)  + \ldots \Big],
\end{equation}
where $L_{\alpha}$, $\alpha=e,\mu,\tau$ are the electron-flavor, muon-flavor, or tau-flavor lepton doublets, $\Lambda$ is the effective energy scale of the operator, and the coefficients $f_{\alpha\beta}=f_{\beta\alpha}$, $\alpha,\beta=e,\mu,\tau$, characterize the operator's distinct flavor components. We define $\Lambda$ such that the largest $f_{\alpha\beta}$ is unity. The amplitude for $0\nu\beta\beta$ is proportional to $f_{ee}$, and the amplitude for $\mu^- \to e^+$ conversion is proportional to $f_{e\mu}$. The series in Eq.~\eqref{definef} also produces rare LNV decays like $K^+ \to \pi^- \mu^+ \mu^+$ and $\tau^- \to \mu^+ \pi^- \pi^-$, as well as lepton-number violation at collider experiments. The limits on $f_{\alpha\beta}/\Lambda$ from these processes are not competitive with limits from $0\nu\beta\beta$ and $\mu^- \to e^+$ conversion for the relevant lepton-flavor structure, and we do not consider them here.\footnote{Recent, detailed discussions and estimates can be found, for instance, in Refs.~\cite{Atre:2005eb,Atre:2009rg,Gluza:2015goa,Gluza:2016qqv,Golling:2016gvc,Quintero:2016iwi}.} The $f_{\alpha\beta}$ do not mix with one another via renormalization-group running due to SM interactions, because lepton-flavor numbers are conserved in the SM.\footnote{We are ignoring the possibility for neutrino Yukawa couplings, which could give rise to lepton-flavor violation at low energies.} We describe the relative strengths of the independent lepton-flavor components of $d\ge7$ operators via coefficients $g_{\alpha\beta\gamma\ldots}$, $\alpha,\beta,\gamma,\ldots=e,\mu,\tau$. These are the analogues of $f_{\alpha\beta}$ in Eq.~\eqref{definef}. In this work, we assume, for simplicity, that the high-scale physics may distinguish between different lepton flavors but treats quark flavors democratically, so we suppress quark-flavor indices. 

The operators listed in Tables~\ref{EstimateTableD5}, \ref{EstimateTableD7}, and \ref{EstimateTableD9} can also be related to Majorana neutrino masses, as discussed in Refs.~\cite{Babu:2001ex,deGouvea:2007qla,Angel:2012ug}.\footnote{The singlet operator $\mathcal{O}_s$ is included in neither of these analyses because one requires very small $\Lambda \sim \mathcal{O}$(GeV) in order to explain the observed neutrino masses. It is, however, discussed briefly in Ref.~\cite{deGouvea:2014lva}.} In a nutshell, the idea is to postulate that UV physics explicitly violates lepton number and that, at the tree level, it manifests itself predominantly as one of the $d\ge 7$ operators listed in Tables~\ref{EstimateTableD5}, \ref{EstimateTableD7}, and \ref{EstimateTableD9}. At the loop level, SM interactions imply that the same physics will lead to nonzero neutrino masses via the Weinberg operator $\mathcal{O}_1$. Hence, these tree-level operators induce operators of lower mass-dimension. Their coefficients can be related by closing external legs into loops and inserting SM interactions. This procedure implies that $f_{\alpha\beta}$ are linear combinations of the $g_{\alpha\beta\ldots}$. 

After electroweak symmetry is broken, the neutrino masses are proportional to the eigenvalues of the matrix $f_{\alpha\beta}$, and the leptonic mixing matrix $U$ is the matrix of its eigenvectors. In Refs.~\cite{deGouvea:2007qla,Angel:2012ug}, the contributions of these operators to the Weinberg operator are estimated using a procedure similar to the one we outline in Sec.~\ref{sec:Estimates}. A range for $\Lambda$ is determined based on the criterion that the largest entries in the neutrino mass matrix lie within $m_{\nu}\in 0.05-0.5$ eV, with higher $\Lambda$ corresponding to lower $m_\nu$. The third column of Tables~\ref{EstimateTableD5}, \ref{EstimateTableD7}, and \ref{EstimateTableD9} lists these ranges of $\Lambda$.

Operators $\mathcal{O}_{4_b}$ in Table~\ref{EstimateTableD7} and $\mathcal{O}_{12_b}$ in Table~\ref{EstimateTableD9} require extra care. These operators are necessarily antisymmetric in the flavors of the two lepton doublets, a feature which was ignored in previous estimates of the contributions of these operators to the neutrino mass matrix. The simplest diagrams one can write down to generate a Majorana neutrino mass for $\mathcal{O}_{4_b}$ ($\mathcal{O}_{12_b}$) are a pair of two-loop (three-loop) diagrams that sum to zero due to this antisymmetry; this is similar to what one encounters in calculating the contributions of neutrino magnetic moment operators to the neutrino mass matrix, as in Ref.~\cite{Bell:2006wi}. Following Ref.~\cite{Bell:2006wi}, the leading contributions to the neutrino mass matrix come from inserting two Yukawa interactions into these diagrams to form either the dimension-seven equivalent of the Weinberg operator or the dimension-five Weinberg operator $\mathcal{O}_1$ at one additional loop level. We update the estimates for the contributions of these operators to the neutrino matrix in Ref.~\cite{deGouvea:2007qla} as follows:
\begin{eqnarray}
\label{O4b-mass}
\mathcal{O}_{4_b}: & & m_{\alpha\beta} = g_{\alpha\beta} \left( \frac{1}{16 \pi^2} + \frac{v^2}{\Lambda^2} \right)\frac{y_t g^2 (y_\beta^2 - y_\alpha^2)}{(16\pi^2)^2} \frac{v^2}{\Lambda}, \\
\mathcal{O}_{12_b}: & & m_{\alpha\beta} = g_{\alpha\beta} \left( \frac{1}{16 \pi^2} + \frac{v^2}{\Lambda^2} \right)\frac{y_t^2 g^2 (y_\beta^2 - y_\alpha^2)}{(16\pi^2)^3} \frac{v^2}{\Lambda}, \label{O12b-mass}
\end{eqnarray}
where $y_t$ is the top-quark Yukawa coupling; $y_\alpha$ is the Yukawa coupling for charged lepton $\alpha = e, \, \mu, \, \tau$; $g$ is the weak coupling constant; and $v$ is the Higgs vacuum expectation value. Because $g_{\alpha\beta}$ is antisymmetric, these matrices have vanishing diagonal elements: $m_{ee} = m_{\mu\mu} = m_{\tau\tau} = 0$. We recalculate the values of $\Lambda$ for each operator such that the largest element of the mass matrix lies within $m_{\nu}\in 0.05-0.5$ eV; the results are listed in Tables~\ref{EstimateTableD7} and \ref{EstimateTableD9}.

It is not possible, in a model-independent way, to relate LNV processes mediated by the new physics, e.g., $0\nu\beta\beta$ and $\mu^- \to e^+$ conversion, because the different $g_{\alpha\beta\ldots}$ are not related. Majorana neutrino masses, however, serve as a link between otherwise disconnected LNV phenomena. If the neutrino masses and the leptonic mixing matrix were known, it would be possible, assuming the physics responsible for nonzero neutrino masses was captured by one of the operators in Tables~\ref{EstimateTableD5}, \ref{EstimateTableD7}, and \ref{EstimateTableD9}, to translate constraints on LNV processes -- like those mentioned below Eq.~\eqref{definef} -- into constraints on other LNV processes.  An important consequence of the connection between Majorana neutrino masses and LNV phenomena is that the observation of any LNV decay, interaction, etc., implies that neutrinos have a Majorana component to their masses, and that the existence of a Majorana neutrino mass implies that some LNV phenomena occur \cite{Schechter:1981cv}. Exactly which processes must occur, however, cannot be predicted a priori. 

\begin{longtable}{|c|c|c||l|}
\caption{Same as Table~\ref{EstimateTableD5}, for the dimension-nine operators featured in this analysis. Naming convention follows from Refs.~\cite{deGouvea:2007qla,Angel:2012ug}, with the exception of the singlet operator $\mathcal{O}_s$ \cite{deGouvea:2014lva}. \label{EstimateTableD9}} \\
\hline
\multirow{2}{*}{$\mathcal{O}$} & \multirow{2}{*}{Operator} & \multirow{2}{*}{$\Lambda$ [TeV]} & $T_{0\nu\beta\beta}$ \\[5pt] \cline{4-4}
& & & $R_{\mu^- e^+}$ \\[5pt] \hline\hline

\multirow{2}{*}{$\mathcal{O}_5$} & \multirow{2}{*}{$(L\overline{H})(LH)(QH)d^c$} & \multirow{2}{*}{$6\times 10^{4-5}$} & $\ln{(2)}\GFU^2 q^2 \frac{\Lambda^2}{Q^{11}}\left[ \GF^2 \frac{1}{q^2}\left(\frac{y_b v^2}{(16\pi^2)^2}\right)^2 + \left( \frac{v}{16\pi^2 \Lambda^2} + \frac{v^3}{\Lambda^4}\right)^2 \right]^{-1} \sim 10^{25}-10^{27}$ yr \\[5pt] \cline{4-4}
& & & $\frac{1}{q}\frac{Q^6}{\Lambda^2} \left[ \GF^2 \frac{1}{q^2}\left(\frac{y_b v^2}{(16\pi^2)^2}\right)^2 + \left( \frac{v}{16\pi^2 \Lambda^2} + \frac{v^3}{\Lambda^4}\right)^2 \right]\sim 10^{-40}-10^{-38}$ \\[5pt] \hline

\multirow{2}{*}{$\mathcal{O}_6$} & \multirow{2}{*}{$(LH)(L\overline{H})(\overline{Q}H)\overline{u^c}$} & \multirow{2}{*}{$2\times 10^{6-7}$} & $\ln{(2)}\GFU^2 q^2 \frac{\Lambda^2}{Q^{11}}\left[ \GF^2 \frac{1}{q^2}\left(\frac{y_t v^2}{(16\pi^2)^2}\right)^2 + \left( \frac{v}{16\pi^2 \Lambda^2} + \frac{v^3}{\Lambda^4}\right)^2 \right]^{-1} \sim 10^{25}-10^{27}$ yr \\[5pt] \cline{4-4}
& & & $\frac{1}{q}\frac{Q^6}{\Lambda^2} \left[ \GF^2 \frac{1}{q^2}\left(\frac{y_t v^2}{(16\pi^2)^2}\right)^2 + \left( \frac{v}{16\pi^2 \Lambda^2} + \frac{v^3}{\Lambda^4}\right)^2 \right]\sim 10^{-37}-10^{-35}$ \\[5pt] \hline

\multirow{2}{*}{$\mathcal{O}_7$} & \multirow{2}{*}{$(LH)(QH)(\overline{Q}H)\overline{e^c}$} & \multirow{2}{*}{$4\times 10^{1-2}$} & $\ln{(2)}\GFU^2 q^2 \frac{\Lambda^2}{Q^{11}} \left[ \GF \frac{v}{(16\pi^2)^2} + \frac{v}{16\pi^2 \Lambda^2} + \frac{v^3}{\Lambda^4} \right]^{-2} \sim 10^{22}-10^{24}$ yr \\[5pt] \cline{4-4}
& & & $\frac{1}{q^2}\frac{Q^6}{\Lambda^2} \left[ \GF \frac{v}{(16\pi^2)^2} + \frac{v}{16\pi^2 \Lambda^2} + \frac{v^3}{\Lambda^4} \right]^2 \sim 10^{-34}-10^{-32}$ \\[5pt] \hline

\multirow{2}{*}{$\mathcal{O}_{9}$} & \multirow{2}{*}{$(LL)(LL)e^c e^c$} & \multirow{2}{*}{$3\times 10^{2-3}$} & $\ln{(2)}\GFU^4 q^4 \left(\frac{16\pi^2}{y_\tau v}\right)^{4} \frac{\Lambda^2}{Q^{11}} \sim 10^{25}-10^{27}$ yr \\[5pt] \cline{4-4}
& & & $\GF^2 \frac{1}{q^4}\left(\frac{y_\tau v}{16\pi^2}\right)^4 \frac{Q^6}{\Lambda^2} \sim 10^{-38}-10^{-36}$ \\[5pt] \hline

\multirow{2}{*}{$\mathcal{O}_{10}$} & \multirow{2}{*}{$(LL)(LQ)e^c d^c$} & \multirow{2}{*}{$6\times 10^{2-3}$} & $\ln{(2)}\GFU^2 q^2 \frac{\Lambda^2}{Q^{11}} \left[ \GF^2 \frac{1}{q^2}\left(\frac{y_\tau y_b v^2}{(16\pi^2)^2}\right)^2 + \left(\frac{y_\tau v}{16\pi^2\Lambda^2}\right)^2 \right]^{-1} \sim 10^{25}-10^{27}$ yr \\[5pt] \cline{4-4}
& & & $\frac{1}{q^2}\frac{Q^6}{\Lambda^2} \left[ \GF^2 \frac{1}{q^2}\left(\frac{y_\tau y_b v^2}{(16\pi^2)^2}\right)^2 + \left(\frac{y_\tau v}{16\pi^2\Lambda^2}\right)^2 \right]\sim 10^{-38}-10^{-36}$ \\[5pt] \hline

\multirow{2}{*}{$\mathcal{O}_{11_a}$} & \multirow{2}{*}{$(LL)(QQ)d^c d^c$} & \multirow{2}{*}{$3-30$} & $\ln{(2)}\GFU^2 q^2 \frac{\Lambda^2}{Q^{11}} \left[\GF^2 \frac{1}{q^2} \left(\frac{y_b^2 g^2 v^2}{(16\pi^2)^3}\right)^2 + \left(\frac{y_b v}{16\pi^2\Lambda^2}\right)^2 \right]^{-1} \sim 10^{22} - 10^{26}$ yr \\[5pt] \cline{4-4}
& & & $\frac{1}{q^2}\frac{Q^6}{\Lambda^2} \left[\GF^2 \frac{1}{q^2} \left(\frac{y_b^2 g^2 v^2}{(16\pi^2)^3}\right)^2 + \left(\frac{y_b v}{16\pi^2\Lambda^2}\right)^2 \right] \sim 10^{-37}-10^{-33}$ \\[5pt] \hline

\multirow{2}{*}{$\mathcal{O}_{11_b}$} & \multirow{2}{*}{$(LQ)(LQ)d^c d^c$} & \multirow{2}{*}{$2\times 10^{3-4}$} & $\ln{(2)}\frac{\Lambda^2}{Q^{11}} \left[ \GF^4 \frac{1}{q^4} \left(\frac{ y_b^2 v^2}{(16\pi^2)^2}\right)^2 + \GF^2 \frac{1}{q^2} \left(\frac{y_b v}{16\pi^2\Lambda^2}\right)^2 + \frac{1}{\Lambda^8} \right]^{-1} \sim 10^{25}-10^{27}$ yr \\[5pt] \cline{4-4}
& & & $\frac{Q^6}{\Lambda^2} \left[ \GF^2 \frac{1}{q^4} \left(\frac{ y_b^2 v^2}{(16\pi^2)^2}\right)^2 + \frac{1}{q^2} \left(\frac{y_b v}{16\pi^2\Lambda^2}\right)^2 + \GFU^2\frac{1}{\Lambda^8} \right] \sim 10^{-38} - 10^{-36}$ \\[5pt] \hline

\multirow{2}{*}{$\mathcal{O}_{12_a}$} & \multirow{2}{*}{$(L\overline{Q})(L\overline{Q})\overline{u^c}\overline{u^c}$} & \multirow{2}{*}{$2\times 10^{6-7}$} & $\ln{(2)}\frac{\Lambda^2}{Q^{11}} \left[ \GF^4 \frac{1}{q^4} \left(\frac{ y_t^2 v^2}{(16\pi^2)^2}\right)^2 + \GF^2 \frac{1}{q^2} \left(\frac{y_t v}{16\pi^2\Lambda^2}\right)^2 + \frac{1}{\Lambda^8} \right]^{-1} \sim 10^{25}-10^{27}$ yr \\[5pt] \cline{4-4}
& & & $\frac{Q^6}{\Lambda^2} \left[ \GF^2 \frac{1}{q^4} \left(\frac{ y_t^2 v^2}{(16\pi^2)^2}\right)^2 + \frac{1}{q^2} \left(\frac{y_t v}{16\pi^2\Lambda^2}\right)^2 + \GFU^2\frac{1}{\Lambda^8} \right] \sim 10^{-38}-10^{-36}$\\[5pt] \hline

\multirow{2}{*}{$\mathcal{O}_{12_b}$} & \multirow{2}{*}{$(LL)(\overline{Q}\overline{Q})\overline{u^c}\overline{u^c}$} & \multirow{2}{*}{$0.3-0.6$} & This operator can not contribute to $0\nu\beta\beta$. \\[5pt] \cline{4-4}
& & & $\frac{1}{q^2}\frac{Q^6}{\Lambda^2} \left[ \GF^2 \left(\frac{v y_t^2 y_d}{(16\pi^2)^2 g^2}\right)^2 + \left(\frac{v y_t}{16\pi^2}\right)^2 \frac{1}{\Lambda^4}\right] \sim 10^{-25} - 10^{-23}$ \\[5pt] \hline

\multirow{2}{*}{$\mathcal{O}_{13}$} & \multirow{2}{*}{$(L\overline{Q})(LL)\overline{u^c}e^c$} & \multirow{2}{*}{$2\times 10^{4-5}$} & $\ln{(2)}\GFU^2 q^2 \frac{\Lambda^2}{Q^{11}} \left[ \GF^2 \frac{1}{q^2} \left(\frac{y_\tau y_t v^2}{(16\pi^2)^2}\right)^2 + \left(\frac{y_\tau v}{16\pi^2\Lambda^2}\right)^2 \right]^{-1} \sim 10^{25}-10^{27}$ yr \\[5pt] \cline{4-4}
& & & $\frac{1}{q^2}\frac{Q^6}{\Lambda^2} \left[ \GF^2 \frac{1}{q^2} \left(\frac{y_\tau y_t v^2}{(16\pi^2)^2}\right)^2 + \left(\frac{y_\tau v}{16\pi^2\Lambda^2}\right)^2 \right] \sim 10^{-37} - 10^{-35}$ \\[5pt] \hline

\multirow{2}{*}{$\mathcal{O}_{14_a}$} & \multirow{2}{*}{$(LL)(Q\overline{Q})\overline{u^c}d^c$} & \multirow{2}{*}{$10^{2-3}$} & $\ln{(2)}\GFU^2 q^2 \frac{\Lambda^2}{Q^{11}} \left[ \GF^2 \frac{1}{q^2} \left(\frac{y_t y_b g^2 v^2}{(16\pi^2)^3}\right)^2 + \left(\frac{y_t v}{16\pi^2\Lambda^2}\right)^2 \right]^{-1} \sim 10^{24}-10^{26}$ yr \\[5pt] \cline{4-4}
& & & $\frac{1}{q^2}\frac{Q^6}{\Lambda^2}\left[ \GF^2 \frac{1}{q^2} \left(\frac{y_t y_b g^2 v^2}{(16\pi^2)^3}\right)^2 + \left(\frac{y_t v}{16\pi^2\Lambda^2}\right)^2 \right] \sim 10^{-37}-10^{-35}$ \\[5pt] \hline

\multirow{2}{*}{$\mathcal{O}_{14_b}$} & \multirow{2}{*}{$(L\overline{Q})(LQ)\overline{u^c}d^c$} & \multirow{2}{*}{$6\times 10^{4-5}$} & $\ln{(2)}\frac{\Lambda^2}{Q^{11}}\left[ \GF^4 \frac{1}{q^4} \left(\frac{y_t y_b v^2}{(16\pi^2)^2}\right)^2 + \GF^2 \frac{1}{q^2} \left(\frac{y_t v}{16\pi^2\Lambda^2}\right)^2 + \frac{1}{\Lambda^8} \right]^{-1} \sim 10^{25}-10^{27}$ yr  \\[5pt] \cline{4-4}
& & & $\frac{Q^6}{\Lambda^2}\left[ \GF^2 \frac{1}{q^4} \left(\frac{y_t y_b v^2}{(16\pi^2)^2}\right)^2 + \frac{1}{q^2} \left(\frac{y_t v}{16\pi^2\Lambda^2}\right)^2 + \GFU^2\frac{1}{\Lambda^8} \right] \sim 10^{-38}-10^{-36}$ \\[5pt] \hline

\multirow{2}{*}{$\mathcal{O}_{15}$} & \multirow{2}{*}{$(LL)(L\overline{L})d^c\overline{u^c}$} & \multirow{2}{*}{$10^{2-3}$} & $\ln{(2)}\GFU^4 \frac{\Lambda^2}{Q^{11}}\left[ \frac{1}{q^2} \left( \frac{y_t y_b g^2 v^2}{(16\pi^2)^3}\right) + \left(\frac{y_t y_b v^2}{(16\pi^2)^2\Lambda^4}\right)\right]^{-2} \sim 10^{24}-10^{26}$ yr \\[5pt] \cline{4-4}
& & & $\GF^2 \frac{Q^6}{\Lambda^2} \left[\frac{1}{q^2} \left( \frac{y_t y_b g^2 v^2}{(16\pi^2)^3}\right) + \left(\frac{y_t y_b v^2}{(16\pi^2)^2\Lambda^4}\right)\right]^{2} \sim 10^{-37}-10^{-35}$ \\[5pt] \hline

\multirow{2}{*}{$\mathcal{O}_{16}$} & \multirow{2}{*}{$(LL)e^c d^c \overline{e^c} \overline{u^c}$} & \multirow{2}{*}{$0.2-2$} & $\ln{(2)}\GFU^2 q^2 \frac{\Lambda^2}{Q^{11}} \left[ \GF^2\frac{1}{q^2}\left(\frac{ y_t y_b g^4 v^2}{(16\pi^2)^4}\right)^2 + \left(\frac{y_\tau v}{16\pi^2\Lambda^2}\right)^2 \right]^{-1} \sim 10^{16}-10^{22}$ yr \\[5pt] \cline{4-4}
& & & $\frac{1}{q^2}\frac{Q^6}{\Lambda^2} \left[ \GF^2\frac{1}{q^2}\left(\frac{ y_t y_b g^4 v^2}{(16\pi^2)^4}\right)^2 + \left(\frac{y_\tau v}{16\pi^2\Lambda^2}\right)^2 \right] \sim 10^{-33} - 10^{-27}$ \\[5pt] \hline

\multirow{2}{*}{$\mathcal{O}_{17}$} & \multirow{2}{*}{$(LL)d^c d^c \overline{d^c} \overline{u^c}$} & \multirow{2}{*}{$0.2-2$} & $\ln{(2)} \GFU^2 q^2 \frac{\Lambda^2}{Q^{11}} \left[ \GF^2\frac{1}{q^2}\left(\frac{y_t y_b g^4 v^2}{(16\pi^2)^4}\right)^2 + \left(\frac{q g^2}{(16\pi^2)^2\Lambda^2}\right)^2 \right]^{-1} \sim 10^{23}-10^{26}$ yr \\[5pt] \cline{4-4}
& & & $\frac{1}{q^2}\frac{Q^6}{\Lambda^2} \left[\GF^2\frac{1}{q^2}\left(\frac{y_t y_b g^4 v^2}{(16\pi^2)^4}\right)^2 + \left(\frac{q g^2}{(16\pi^2)^2\Lambda^2}\right)^2 \right] \sim 10^{-37} - 10^{-34}$ \\[5pt] \hline

\multirow{2}{*}{$\mathcal{O}_{18}$} & \multirow{2}{*}{$(LL)d^c u^c \overline{u^c} \overline{u^c}$} & \multirow{2}{*}{$0.2-2$} & $\ln{(2)} \GFU^2 q^2 \frac{\Lambda^2}{Q^{11}} \left[ \GF^2\frac{1}{q^2}\left(\frac{y_t y_b g^4 v^2}{(16\pi^2)^3}\right)^2 + \left(\frac{q g^2}{(16\pi^2)^2\Lambda^2}\right)^2 \right]^{-1} \sim 10^{23}-10^{26}$ yr \\[5pt] \cline{4-4}
& & & $\frac{1}{q^2}\frac{Q^6}{\Lambda^2} \left[\GF^2\frac{1}{q^2}\left(\frac{y_t y_b g^4 v^2}{(16\pi^2)^3}\right)^2 + \left(\frac{q g^2}{(16\pi^2)^2\Lambda^2}\right)^2 \right] \sim 10^{-37} - 10^{-34}$ \\[5pt] \hline

\multirow{2}{*}{$\mathcal{O}_{19}$} & \multirow{2}{*}{$(LQ) d^c d^c \overline{e^c} \overline{u^c}$} & \multirow{2}{*}{$0.1-1$} & $\ln{(2)}\frac{\Lambda^2}{Q^{11}} \left[ \GF^4 \frac{1}{q^4} \left(\frac{ y_t y_b^2 y_e v^2}{(16\pi^2)^3}\right)^2 + \GF^2 \frac{1}{q^2} \left(\frac{y_b v}{16\pi^2\Lambda^2}\right)^2 + \frac{1}{\Lambda^8}\right]^{-1} \sim 10^{10}-10^{19}$ yr \\[5pt] \cline{4-4}
& & & $\frac{Q^6}{\Lambda^2}\left[ \GF^2 \frac{1}{q^4} \left(\frac{ y_t y_b^2 y_\mu v^2}{(16\pi^2)^3}\right)^2 + \frac{1}{q^2} \left(\frac{y_b v}{16\pi^2\Lambda^2}\right)^2 + \GFU^2\frac{1}{\Lambda^8}\right] \sim 10^{-30}-10^{-21}$ \\[5pt] \hline

\multirow{2}{*}{$\mathcal{O}_{20}$} & \multirow{2}{*}{$(L\overline{Q}) d^c \overline{u^c} \overline{e^c} \overline{u^c}$} & \multirow{2}{*}{$4-40$} & $\ln{(2)}\frac{\Lambda^2}{Q^{11}}\left[ \GF^4 \frac{1}{q^4} \left(\frac{y_t^2 y_b y_e v^2}{(16\pi^2)^3}\right)^2 + \GF^2 \frac{1}{q^2}\left(\frac{y_t v}{16\pi^2\Lambda^2}\right)^2 + \frac{1}{\Lambda^8}\right]^{-1} \sim 10^{19}-10^{25}$ yr \\[5pt] \cline{4-4}
& & & $\frac{Q^6}{\Lambda^2} \left[ \GF^2 \frac{1}{q^4} \left(\frac{y_t^2 y_b y_\mu v^2}{(16\pi^2)^3}\right)^2 + \frac{1}{q^2}\left(\frac{y_t v}{16\pi^2\Lambda^2}\right)^2 + \GFU^2\frac{1}{\Lambda^8}\right] \sim 10^{-35}-10^{-30}$ \\[5pt] \hline

\multirow{2}{*}{$\mathcal{O}_s$} & \multirow{2}{*}{$e^c e^c u^c u^c \overline{d^c} \overline{d^c}$} & \multirow{2}{*}{$10^{-3}$} & $\ln{(2)}\frac{\Lambda^2}{Q^{11}}\left[ \GF^4 \frac{1}{q^4} \left(\frac{ y_t^2 y_b^2 y_e^2 v^2}{(16\pi^2)^4}\right)^2 + \GF^2 \frac{1}{q^2} \left(\frac{y_t y_b y_e v}{(16\pi^2)^2\Lambda^2}\right)^2  + \frac{1}{\Lambda^8}\right]^{-1} \sim 10^{-20}-10^{-10}$ yr \\[5pt] \cline{4-4}
& & & $\frac{Q^6}{\Lambda^2}\left[ \GF^2 \frac{1}{q^4} \left(\frac{ y_t^2 y_b^2 y_\mu^2 v^2}{(16\pi^2)^4}\right)^2 + \frac{1}{q^2} \left(\frac{y_t y_b y_\mu v}{(16\pi^2)^2\Lambda^2}\right)^2  + \GFU^2\frac{1}{\Lambda^8}\right] \sim 10^{-1}-10^{9}$ \\ \hline
\end{longtable}

Even partial information on neutrino masses and lepton mixing allows one to relate different LNV phenomena. As an example, we discuss the connection between Majorana neutrino masses and LNV phenomena using $0\nu\beta\beta$ and $\mu^- \to e^+$ conversion assuming the Weinberg operator $\mathcal{O}_1$ captures the bulk of LNV phenomena. If neutrino exchange dominates these processes -- the case of $\mathcal{O}_1$ -- the rate of $0\nu\beta\beta$ is proportional to
\begin{equation}
\label{definemee}
|m_{ee}|^2 \equiv \left| U_{e1}^2 m_1 + U_{e2}^2 m_2 e^{i \alpha_1} + U_{e3}^2 m_3 e^{i \alpha_2} \right|^2,
\end{equation}
while the rate of $\mu^- \to e^+$ conversion is proportional to
\begin{equation}
\label{definememu}
|m_{e\mu}|^2 \equiv \left|  U_{e1} U_{\mu 1} m_1 + U_{e2} U_{\mu 2} m_2 e^{i \alpha_1} + U_{e3} U_{\mu 3} m_3 e^{i \alpha_2} \right|^2,
\end{equation}
where $m_i$ is the mass of $\nu_i$, $U_{\alpha i}$ are the elements of the leptonic mixing matrix $U$, and $\alpha_i$ are potential Majorana phases. If nothing were known about the neutrino masses and mixing parameters, nothing could be said about $m_{ee}$ in relation to $m_{e\mu}$. However, from current measurements of the leptonic mixing matrix and the neutrino mass-squared differences \cite{Agashe:2014kda}, we find that $m_{ee}$ and $m_{e\mu}$ cannot simultaneously vanish, for any value of the unknown $m_1$, $\alpha_1$ and $\alpha_2$ parameters. This implies that if LNV manifests itself predominantly via $\mathcal{O}_1$  at least one of $0\nu\beta\beta$ and $\mu^- \to e^+$ conversion must occur. 

Neutrino exchange does not, however, always dominate the amplitudes for these processes, as we discuss in detail in Sec.~\ref{sec:Estimates}. Even so, we have verified, for all operators in Tables~\ref{EstimateTableD5}, \ref{EstimateTableD7}, and \ref{EstimateTableD9}, that if $m_{ee}$ ($m_{e\mu}$) is nonzero the amplitude for $0\nu\beta\beta$ ($\mu^- \to e^+$ conversion) does not vanish  as long as the dominant LNV physics is captured by one of the operators in Tables~\ref{EstimateTableD5}, \ref{EstimateTableD7}, and \ref{EstimateTableD9}. There is no guarantee, of course, that the nonzero rate is within experimental reach. If more operators are present with commensurate strength, we cannot rule out the possibility of fortuitous cancellations. 


\section{Estimates and Comparisons}
\label{sec:Estimates}

In this section, we describe the process used for estimating $0\nu\beta\beta$ half-lives ($\TBB$) and $\mu^-\to e^+$ conversion rates ($\RME$), concentrating, for concreteness, on $\mathcal{O}_{14_b}$. In Section~\ref{subsec:BBEstimate}, we discuss $0\nu\beta\beta$, and in Section~\ref{subsec:MuEEstimate} we discuss $\mu^-\to e^+$ conversion in nuclei. We estimate the values of diagrams with incoming (outgoing) down quarks and outgoing (incoming) up quarks for $0\nu\beta\beta$ ($\mu^-\to e^+$ conversion), and we bypass effects from hadronic currents, nuclear matrix elements, phase-space integration, etc.. In order to make comparisons with existing and future experimental results, we take advantage of existing bounds on \cite{KamLAND-Zen:2016pfg} or calculations of \cite{Simkovic:2000ma,Domin:2004tk} the light neutrino exchange contribution, as will become clear momentarily.  

\subsection{Neutrinoless Double Beta Decay}
\label{subsec:BBEstimate}
Here, we discuss how we estimate $\TBB$ for the operator $\mathcal{O}_{14_b} = (L\overline{Q})(LQ)\overline{u^c}d^c$. We separate the discussion into contributions at tree level, one loop, and two loops. We reemphasize that these are rough estimates aimed at capturing the dominant contributing factors to $0\nu\beta\beta$ and comparing these different contributions. Much more thorough calculations involving hadronic currents, etc., are necessary in order to extract accurate bounds. For our purposes, however, order-of-magnitude estimates are sufficient. 
\begin{figure}[!htbp]
\begin{center}
\subfigure[]{\includegraphics[width=2.3in]{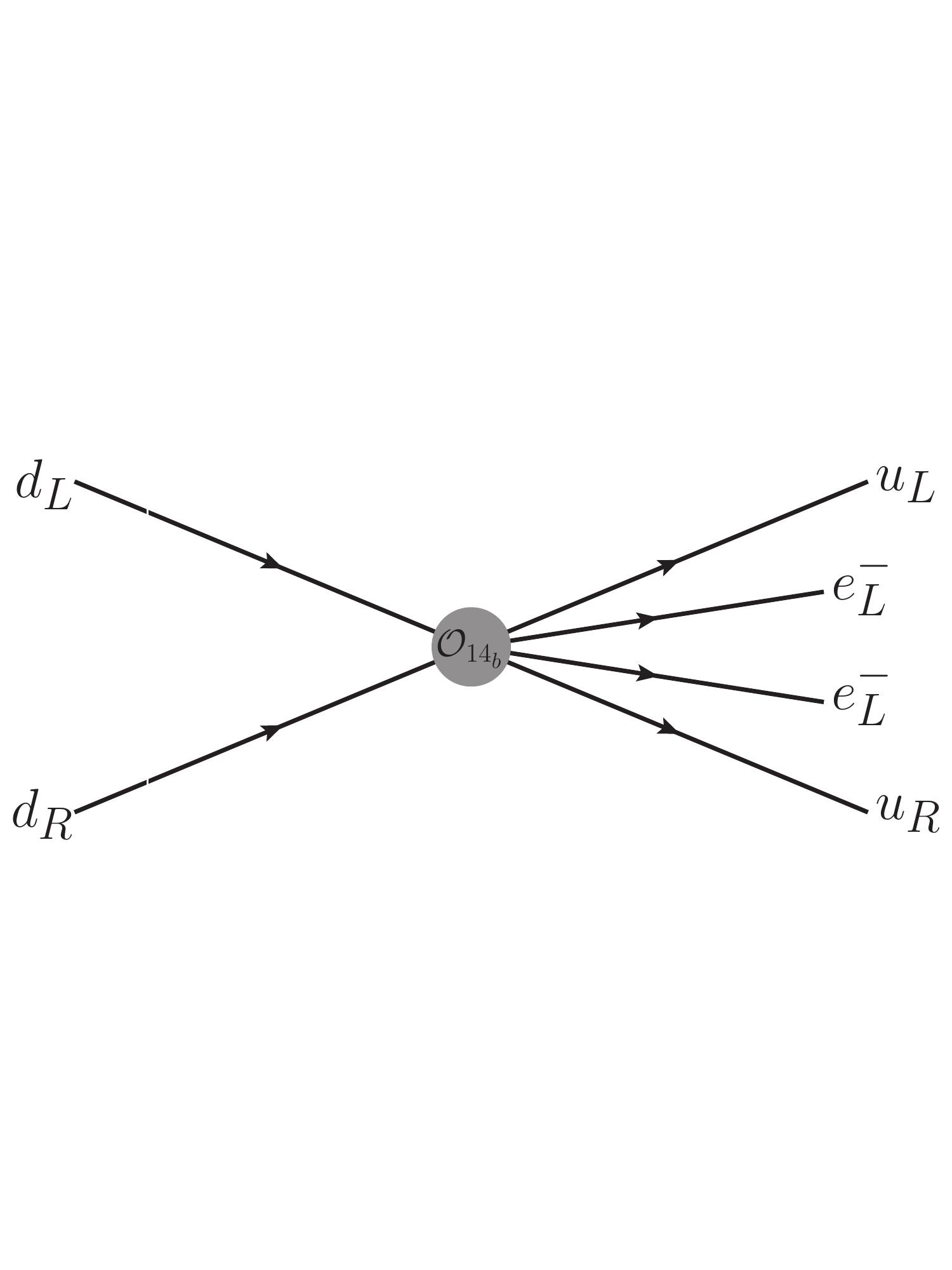}}
\subfigure[]{\includegraphics[width=2.3in]{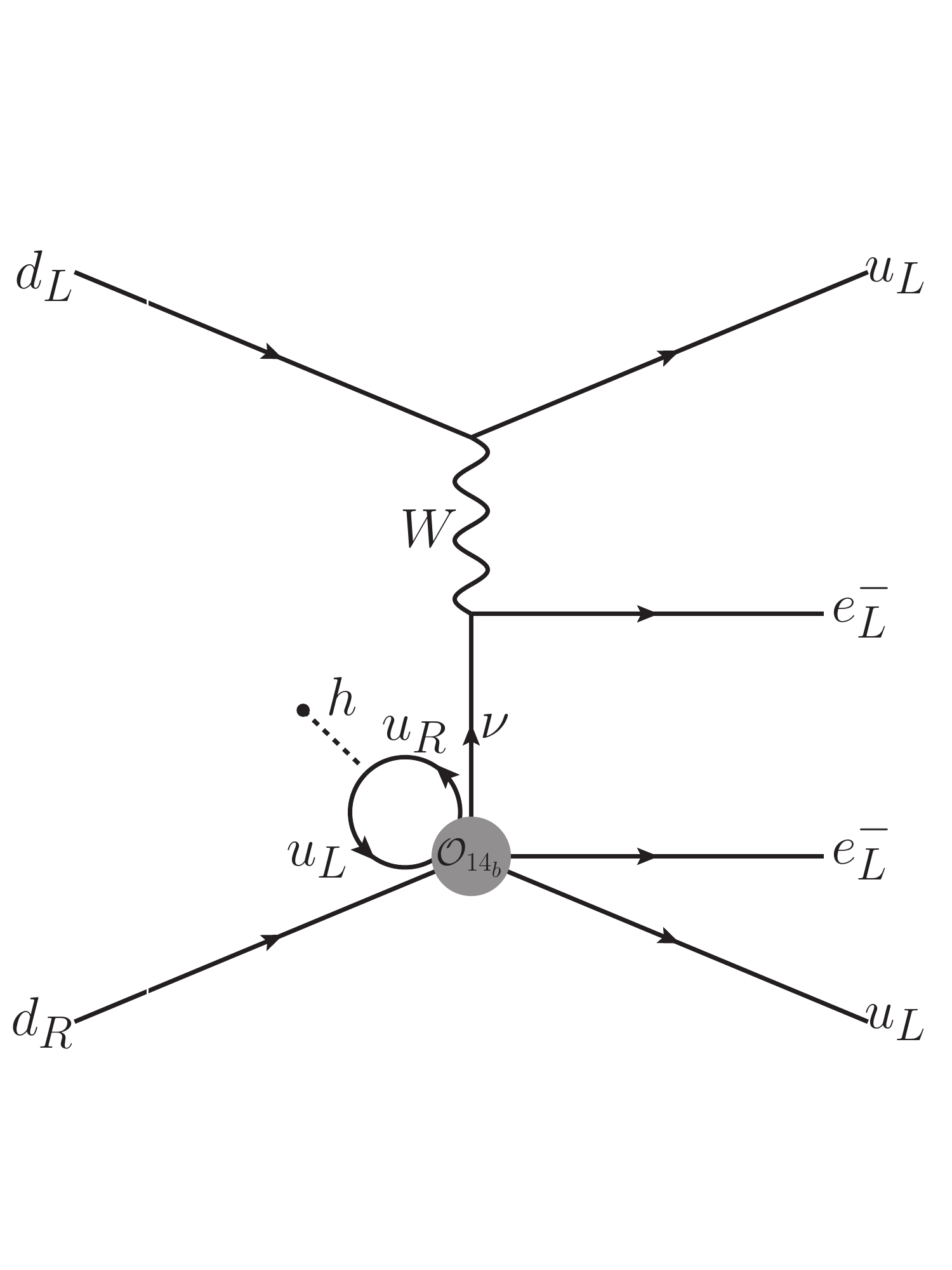}}
\subfigure[]{\includegraphics[width=2.3in]{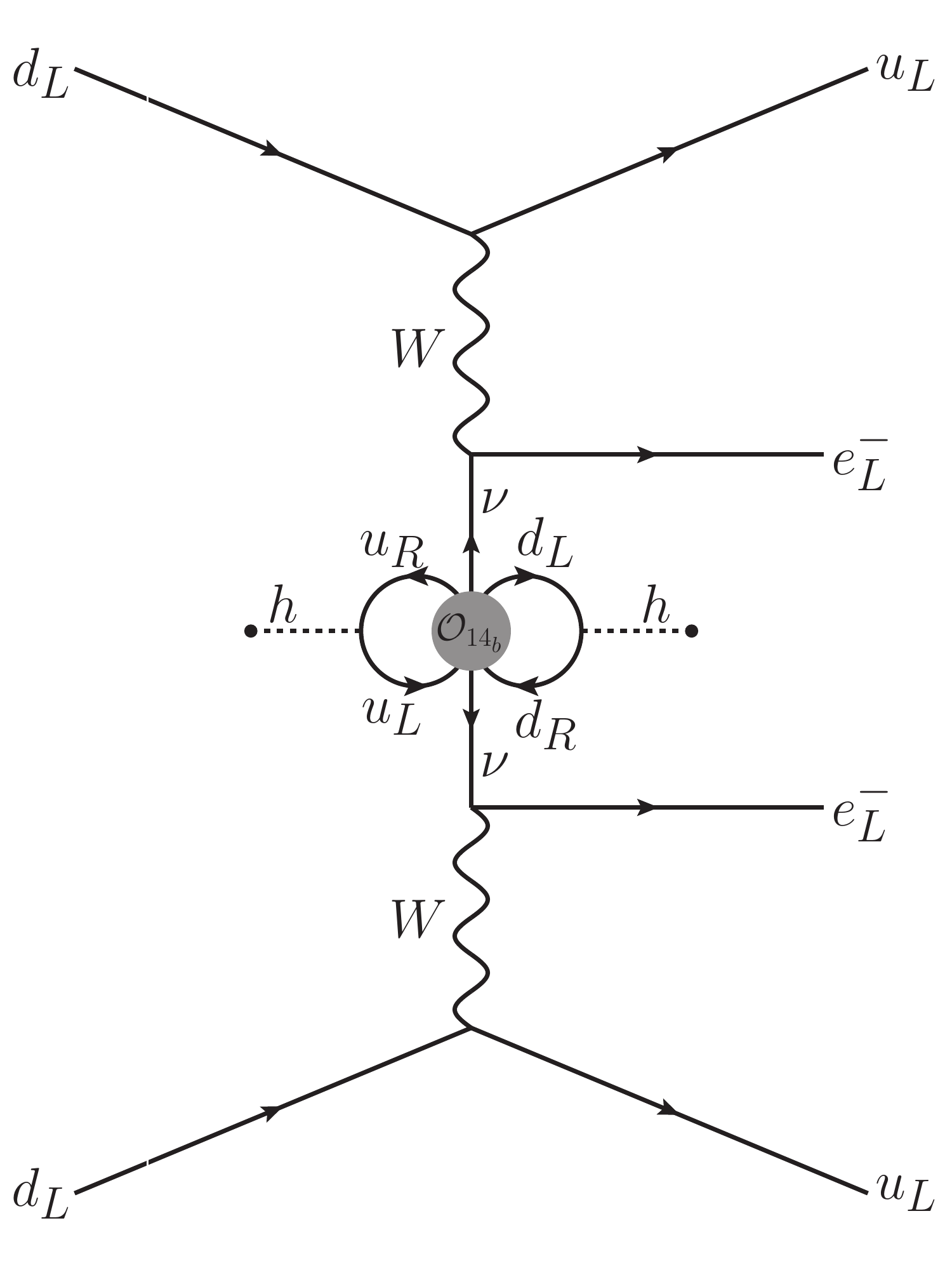}}
\end{center}
\caption{Feynman diagrams contributing to $0\nu\beta\beta$ from the operator $\mathcal{O}_{14_b} = (L\overline{Q})(LQ)\overline{u^c}d^c$. The dominant contributions scaling as $\Gamma \sim \Lambda^{-10}$ (a), $\Lambda^{-6}$ (b), and $\Lambda^{-2}$ (c) are depicted.}
\label{fig:Feynman}
\end{figure}

\subsubsection{Tree level}
Fig.~\ref{fig:Feynman}(a) depicts the dominant tree-level contribution from $\mathcal{O}_{14_a}$ for $0\nu\beta\beta$. The amplitude scales as $\Lambda^{-5}$ since $\mathcal{O}_{14_a}$ has mass-dimension nine. We use the variable $Q$, which has dimensions of mass and encodes all information related to phase-space, nuclear matrix elements, etc., in order to convert the diagram into a decay rate, via naive dimensional analysis. $Q$ is naively of order the $Q$-value of the decay process, a few MeV. Our estimate is
\begin{equation}
\Gamma_{0\nu\beta\beta}^{(0)} = |g_{ee}|^2 \frac{Q^{11}}{\Lambda^{10}},
\label{eq:TBBTree}
\end{equation}
where $g_{ee}$ reflects the fact that this contribution requires both of the lepton doublets to be of electron flavor. 

\subsubsection{One loop}
Here we consider the diagram shown in Fig.~\ref{fig:Feynman}(b). When calculating a loop contribution, we assume the momentum cutoff scale to be $\Lambda$, above which the effective field theory approach is no longer valid. Each loop also contributes a factor of $(16\pi^2)^{-1}$ to the amplitude. We estimate the contribution of this loop to the amplitude to be
\begin{equation}
\int \frac{d^4 p}{(2\pi)^4p^2} \sim \frac{\Lambda^2}{16\pi^2}.
\end{equation}
The amplitude, therefore, scales as $\Lambda^{-3}$. The Higgs boson can be replaced by its vacuum expectation value $v$ which multiplies its coupling to the up-type quark in the loop. We choose to take all effective operators to be quark-flavor universal, so the largest contribution to this diagram comes from the top quark, proportional to $y_t$, the top quark Yukawa coupling. The $W-$boson propagator and couplings contribute a factor of $G_F/\sqrt{2}$. The dominant contribution from the neutrino propagator scales like $1/q^2$, which we estimate is of order $(100~\rm MeV)^{-2}$, the typical distance scale between nucleons. Therefore, we estimate
\begin{equation}
\Gamma_{0\nu\beta\beta}^{(1)} = |g_{ee}|^2 \GF^2 \left(\frac{1}{q^2} \right) \left( \frac{v y_t}{16\pi^2} \right)^2 \frac{Q^{11}}{\Lambda^6}.
\label{eq:TBBOneLoop}
\end{equation}
Since the neutrino propagator is not exactly point-like, the phase-space-matrix-element-etc.-$Q^2$ factor here is not identical to the one in Eq.~(\ref{eq:TBBTree}). The difference -- not more than an order of magnitude -- is too small to impact our results and will be ignored.

\subsubsection{Two loop}
The dominant contribution at two-loop order comes from the diagram shown in Fig.~\ref{fig:Feynman}(c). Here, one loop contributes the same factor discussed above, and the second contributes the same factor but with the bottom quark Yukawa coupling $y_b$ instead of the top quark Yukawa coupling $y_t$. Additionally, there are two $W-$boson propagators instead of one. The neutrino propagator contributes a factor proportional to $1/q^2$ on top of the mass-insertion associated to the two-loop diagram. The estimate, therefore, is
\begin{equation}
\Gamma_{0\nu\beta\beta}^{(2)} = |g_{ee}|^2 \GF^4 \left(\frac{1}{q^2}\right)^2 \left(\frac{y_t y_b v^2}{(16\pi^2)^2}\right)^2 \frac{Q^{11}}{\Lambda^2} = \GF^4 \left(\frac{1}{q^2}\right)^2 |m_{ee}|^2 Q^{11}.
\label{eq:TBBTwoLoop}
\end{equation}
This diagram is exactly the neutrino exchange process discussed in Section~\ref{sec:theory}, hence we have rewritten the width as proportional to $|m_{ee}|^2$. For $\mathcal{O}_{14_b}$, the neutrino mass matrix in the flavor basis is estimated to be
\begin{equation}
m_{\alpha\beta} = \frac{g_{\alpha\beta}}{\Lambda}\frac{y_t y_b v^2}{(16\pi^2)^2},
\label{eq:m_ab}
\end{equation}
$\alpha,\beta=e,\mu,\tau$. As in the case of Eq.~(\ref{eq:TBBOneLoop}), the phase-space-matrix-element-etc.-$Q^2$ factor here is not identical to the one in Eq.~(\ref{eq:TBBTree}) but the difference can, given our goals, be safely ignored. 

We use the results from the KamLAND-Zen experiment, along with the upper bound they compute for $|m_{ee}|$, in order to extract the value of $Q^{11}$ by requiring that Eq.~(\ref{eq:TBBTwoLoop}) exactly reproduces the KamLAND-Zen result, i.e., we obtain the lower bound on the half-life for the quoted upper bound on $|m_{ee}|$. Concretely, the bound $\TBB>1.07\times 10^{26}$ (90\% CL) from KamLAND-Zen, which can be translated into $m_{ee}<100$ meV -- here we make a concrete choice about the relevant nuclear matrix element -- results into $Q=11$~MeV.

\begin{figure}[!htbp]
\centering
\includegraphics[width=0.8\linewidth]{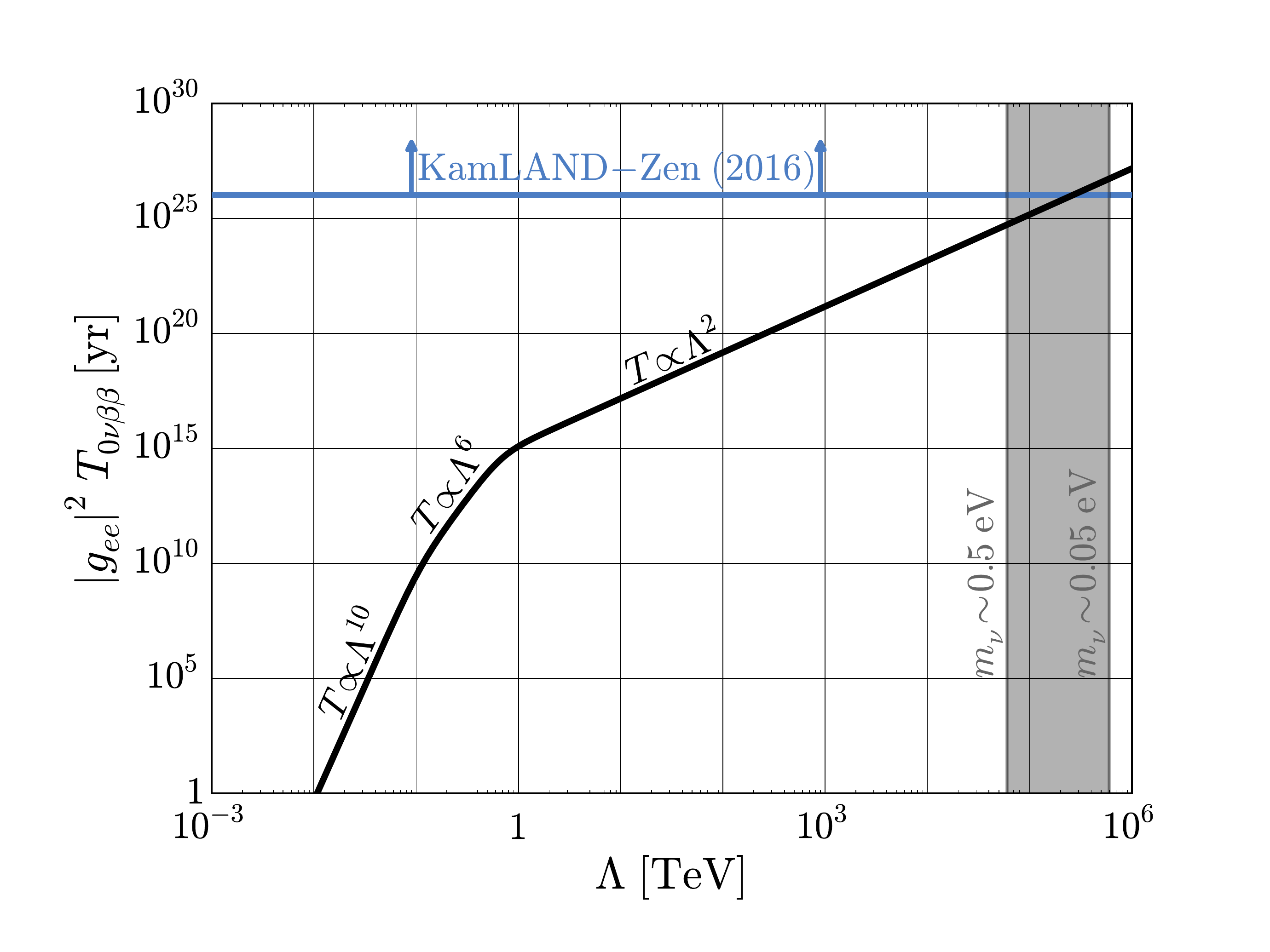}
\caption{The $0\nu\beta\beta$ half-life $\TBB$ as a function of the scale of new physics $\Lambda$ for operator $\mathcal{O}_{14_b}$. The pink line displays the current bound (assuming $|g_{ee}|^2 = 1$) by the KamLAND-Zen experiment~\cite{KamLAND-Zen:2016pfg}, $\TBB > 1.07\times 10^{26}$ yr (90\% CL), and the grey region shows the range of $\Lambda$ necessary to generate neutrino masses between $0.05$ and $0.5$ eV. Three distinct regions are visible on the graph, where $T\propto \Lambda^{10}$, $\Lambda^{6}$, and $\Lambda^{2}$. These regions correspond to when the diagrams in Fig.~\ref{fig:Feynman}(a), (b), and (c) are dominant in this process, respectively.}
\label{fig:TvsL}
\end{figure}

For $\mathcal{O}_{14_b}$, the tree-level, one-loop, and two-loop processes add incoherently, so $\Gamma_{0\nu\beta\beta} \equiv \Gamma_{0\nu\beta\beta}^{(0)} + \Gamma_{0\nu\beta\beta}^{(1)} + \Gamma_{0\nu\beta\beta}^{(2)}$. Fig.~\ref{fig:TvsL} depicts the half-life $\TBB = \log{(2)}/\Gamma_{0\nu\beta\beta}$ as a function of $\Lambda$. Also shown are the current bound on $\TBB > 1.07\times 10^{26}$ yr (90\% CL) from the KamLAND-Zen experiment~\cite{KamLAND-Zen:2016pfg} along with the range of $\Lambda$ where $\mathcal{O}_{14_b}$ leads to neutrino masses between $0.05$ and $0.5$ eV, as listed in Tables~\ref{EstimateTableD9}. Generically, a subset of the tree-level, one-loop and two-loop diagrams may interfere with one another for a given operator. If so, then the transitions between different $\Lambda$-dependencies in Fig.~\ref{fig:TvsL} will be smoothed out. If we assume $|g_{ee}|^2 = 1$ and that $\mathcal{O}_{14_b}$ is responsible for neutrino masses, we estimate $\TBB \sim 10^{25}-10^{27}$ years. On the other hand, assuming $|g_{ee}|^2 = 1$, the current bounds on $\TBB$ translates into $\Lambda \gtrsim 10^{5}$ TeV. The current upper bound on $\TBB$ implies that the dominant contribution to $0\nu\beta\beta$ coming from UV physics that manifests itself at the tree-level as $\mathcal{O}_{14_b}$ comes from massive neutrino exchange. 

\subsection{$\mu^-\to e^+$ conversion}
\label{subsec:MuEEstimate}
In order to estimate the rate of $\mu^-\to e^+$ conversion, we first address the muon capture rate. As this is a weak-interaction process, it is proportional to the probability density function of the incoming muon $|\psi_{100}(0)|^2$ (which we assume to be in the $1$s ground state of the atom), so we estimate
\begin{equation}
\Gamma (\mu\ \text{capture}) \sim \GF^2 Q^2 \left( \frac{Z_\text{eff}^3}{\pi (a_0 m_e/m_\mu)^3)}\right),
\end{equation}
where $a_0$ is the Bohr radius and $Q$ is a number with dimensions of mass that contains information regarding phase-space, nuclear matrix elements, etc., similar to the equivalent variable in the $0\nu\beta\beta$-decay discussion. Note that, here, the $Q$-value of the reaction is of order the muon mass. The last factor in parenthesis is $|\psi_{100}(0)|^2$. The rate for $\mu^- \to e^+$ conversion depends on $|\psi_{100}(0)|^2$ as well, which cancels out in estimating $\RME$ 

\begin{figure}[!htbp]
\begin{center}
\subfigure[]{\includegraphics[width=2.3in]{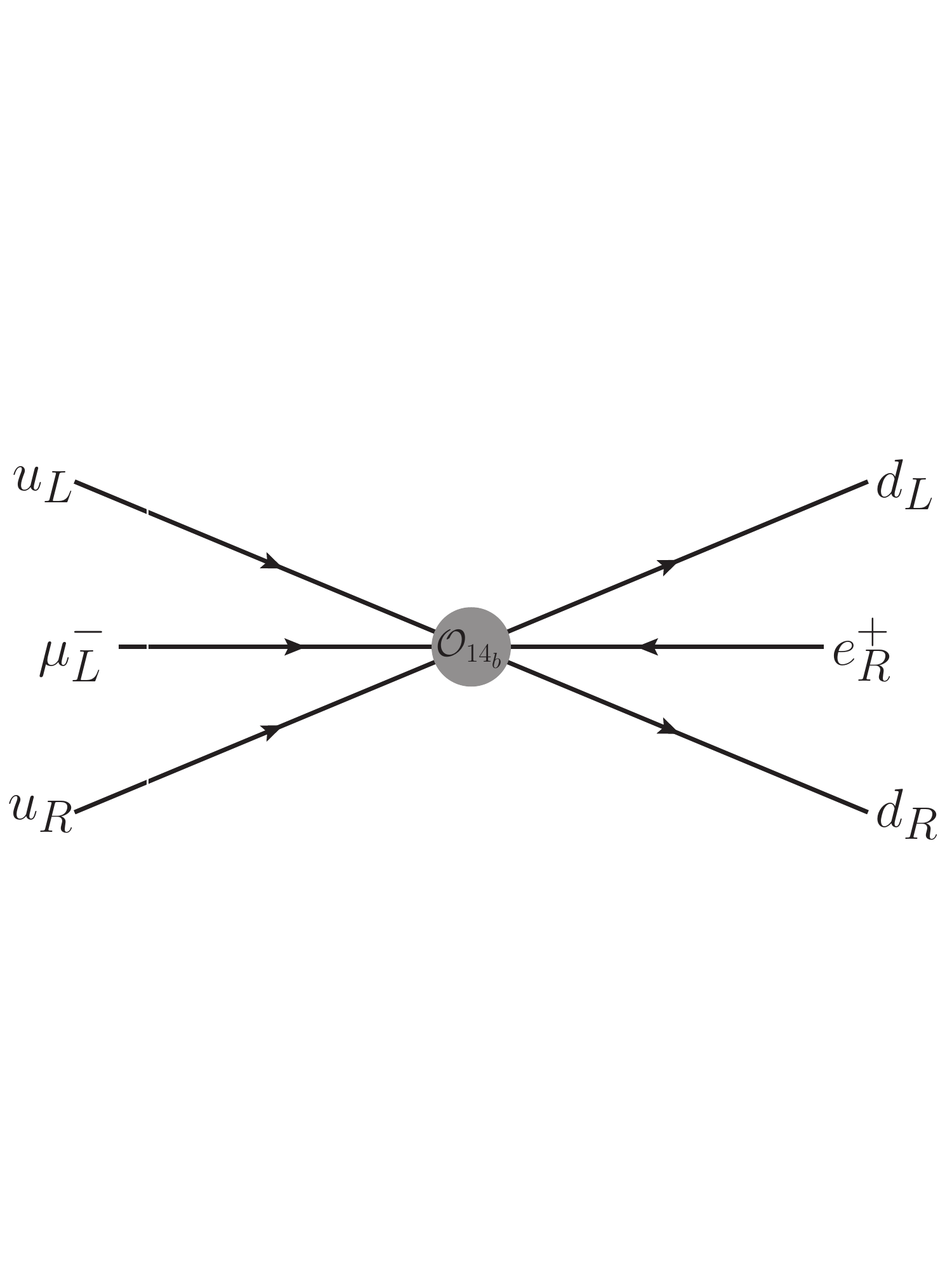}}
\subfigure[]{\includegraphics[width=2.3in]{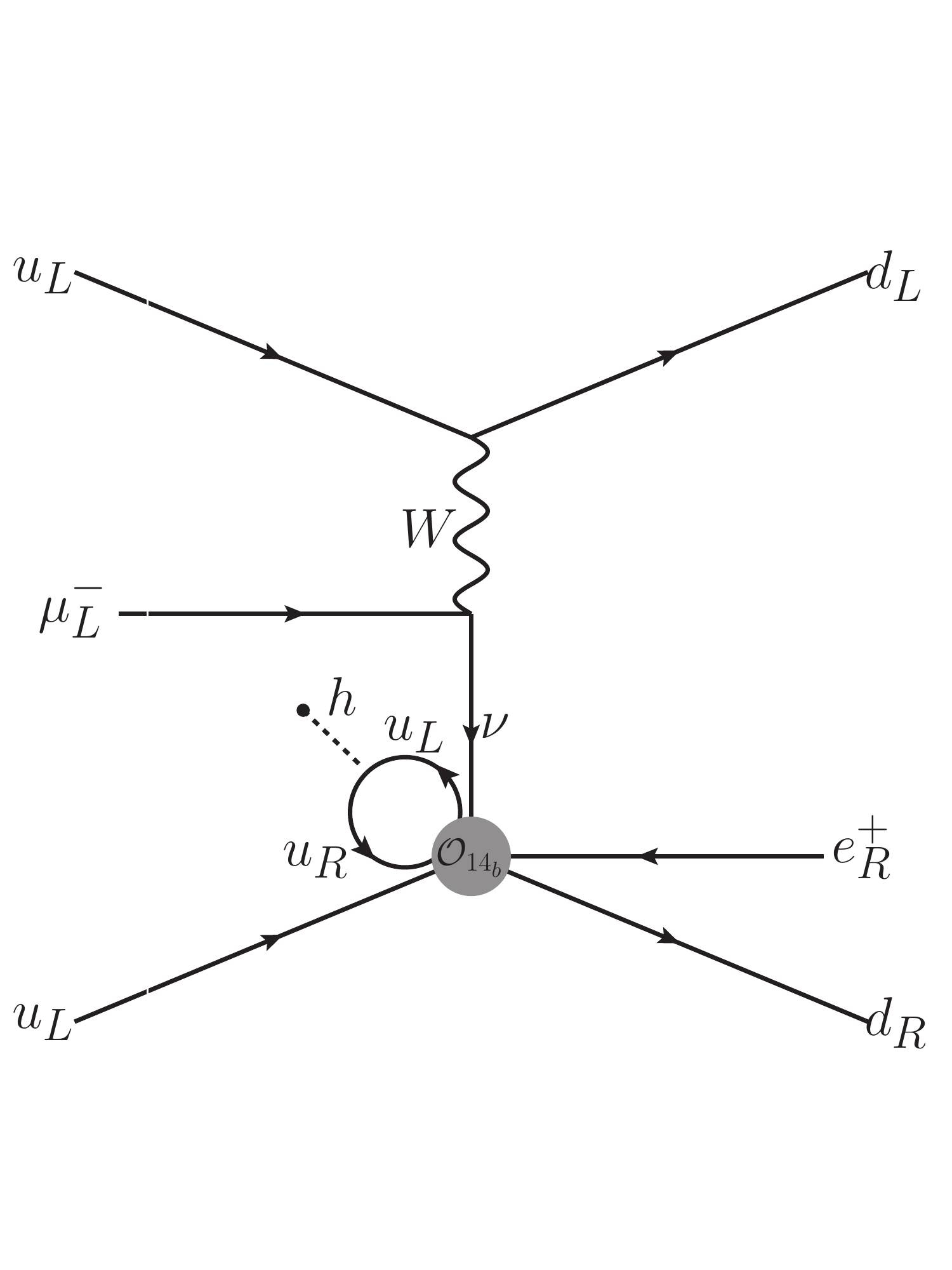}}
\subfigure[]{\includegraphics[width=2.3in]{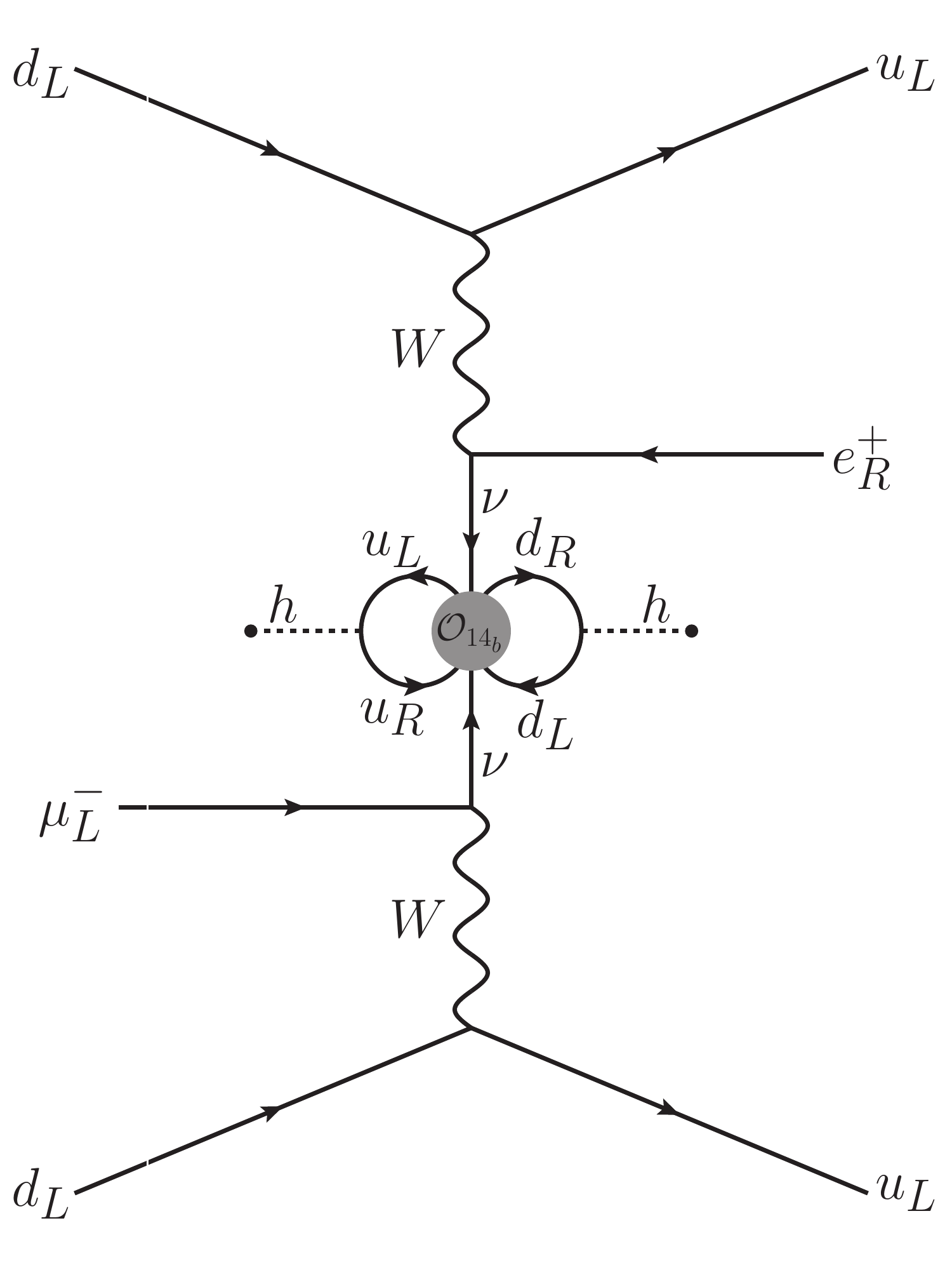}}
\end{center}
\caption{Feynman diagrams contributing to $\mu^-\to e^+$ conversion from the operator $\mathcal{O}_{14_b} = (L\overline{Q})(LQ)\overline{u^c}d^c$. The dominant contributions scaling as $\Gamma \sim \Lambda^{-10}$ (a), $\Lambda^{-6}$ (b), and $\Lambda^{-2}$ (c) are shown.}
\label{fig:MuToE}
\end{figure}
Fig.~\ref{fig:MuToE} shows the dominant diagrams contributing at tree-level (a), one loop (b), and two loops (c) to $\mu^-\to e^+$ conversion. These are very similar to Figs.~\ref{fig:Feynman}(a), (b), and (c), respectively. The contributions to $\RME$ from Figs.~\ref{fig:MuToE}(a), (b), and (c) can be estimated following the same steps that led to Eqs.~(\ref{eq:TBBTree}),~(\ref{eq:TBBOneLoop}), and~(\ref{eq:TBBTwoLoop}), respectively. We find
\begin{align}
\RME^{(0)} &= |g_{e\mu}|^2 \GFU^2\frac{Q^6}{\Lambda^{10}}, \\
\RME^{(1)} &= |g_{e\mu}|^2 \left(\frac{1}{q^2}\right) \left(\frac{y_t v}{16\pi^2}\right)^2 \frac{Q^6}{\Lambda^{6}}, \\
\RME^{(2)} &= |g_{e\mu}|^2 \left(\frac{G_F}{\sqrt{2}}\right)^2 \left(\frac{1}{q^2}\right)^2 \left(\frac{y_t y_b v^2}{(16\pi^2)^2}\right)^2 \frac{Q^6}{\Lambda^2}. \label{eq:RMETwoLoop}
\end{align}
Similar to Eq.~(\ref{eq:TBBTwoLoop}), Eq.~(\ref{eq:RMETwoLoop}) can be written as a coefficient times $|m_{e\mu}|^2$, see Eq.~(\ref{eq:m_ab}). As in Sec.~\ref{subsec:BBEstimate}, the $Q^2$ factors are not strictly the same for the tree-level, one-loop, and two-loop contributions, but we assume that differences are sufficiently small and can be safely ignored. 
 
Refs.~\cite{Simkovic:2000ma,Domin:2004tk} estimated $\RME$ for light neutrino exchange,
\begin{equation}
\RME = \left(2.6\times 10^{-22}\right)|{\cal M}_{e\mu^+}|^2\frac{|m_{e\mu}|^2}{m_e^2},
\label{eq:calc}
\end{equation}
where $m_e$ is the electron mass and $|{\cal M}_{e\mu^+}|$ is the nuclear matrix element, estimated to lie, for titanium, between 0.03 and 0.5. Similar to what we did in Sec.~\ref{subsec:BBEstimate}, we solve for $Q$ in the estimates above so that Eq.~(\ref{eq:RMETwoLoop}) matches the more precise estimate, Eq.~(\ref{eq:calc}), for $|{\cal M}_{e\mu^+}|=0.1$, which we assume is the value of the nuclear matrix element for aluminum to sodium transition.

\begin{figure}[!htbp]
\centering
\includegraphics[width=0.8\linewidth]{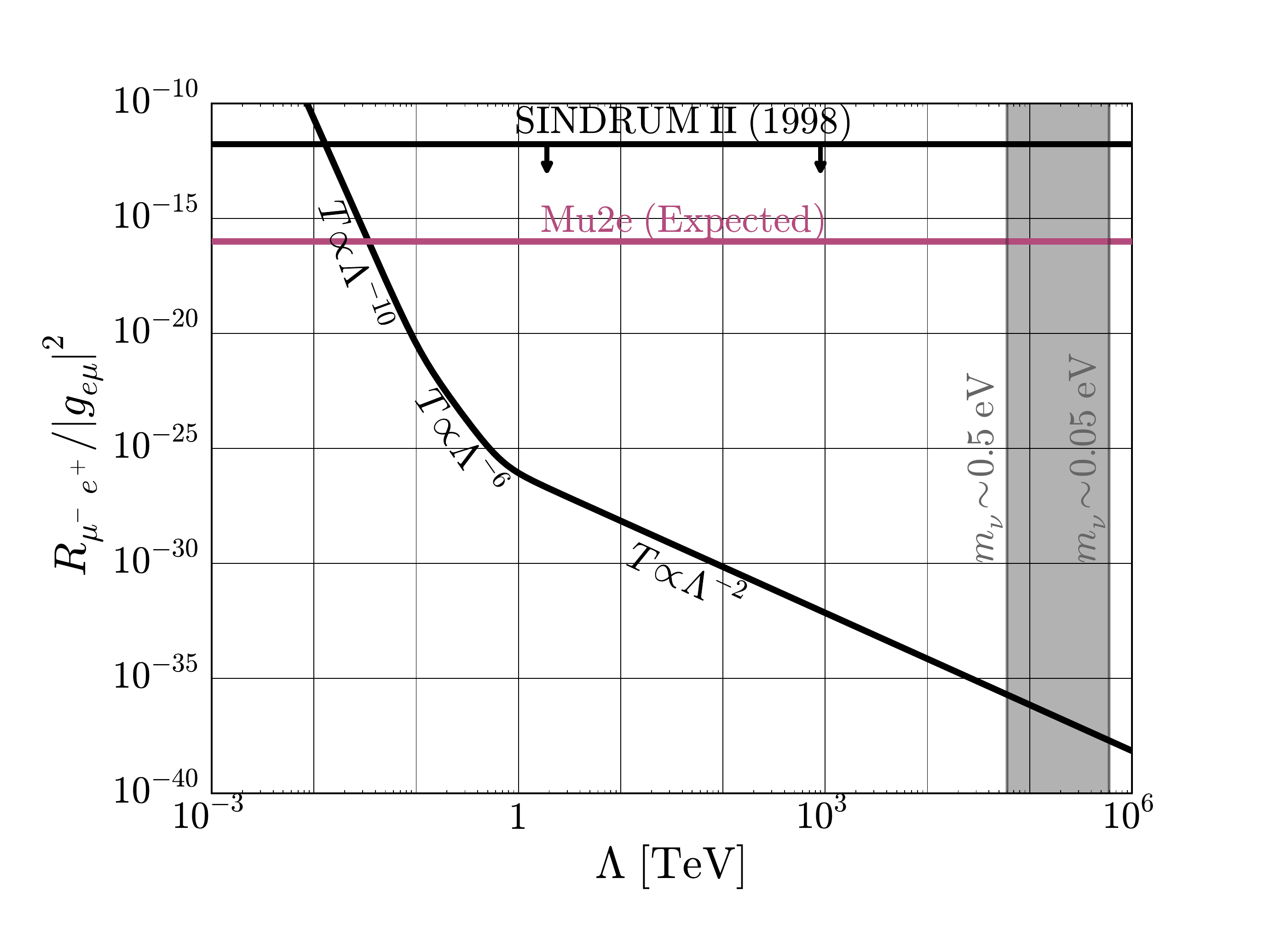}
\caption{The $\mu^-\to e^+$ conversion rate $\RME$ as a function of the scale of new physics $\Lambda$ for operator $\mathcal{O}_{14_b}$. The black line displays the current bound by the SINDRUM II Collaboration, while the blue line indicates our estimate of the reach of the Mu2e experiment, both assuming $|g_{e\mu}|^2 = 1$. The grey region highlights the range of $\Lambda$ necessary to generate neutrino masses between $0.05$ and $0.5$ eV. Three distinct regions are visible on the graph, where $T\propto \Lambda^{10}$, $\Lambda^{6}$, and $\Lambda^{2}$. These regions correspond to when the diagrams in Fig.~\ref{fig:Feynman}(a), (b), and (c) are dominant in this process, respectively.}
\label{fig:RvsL}
\end{figure}

As with $0\nu\beta\beta$, these diagrams add incoherently, so $\RME = \RME^{(0)} + \RME^{(1)} +\RME^{(2)}$. Fig.~\ref{fig:RvsL} depicts the normalized conversion rate $\RME$ as a function of $\Lambda$. Also shown is the range of $\Lambda$ where $\mathcal{O}_{14_b}$ leads to neutrino masses between $0.05$ and $0.5$ eV, as listed in Table~\ref{EstimateTableD9}. As before, interference between the tree-level, one-loop and two-loop diagrams would smooth out the transitions between different $\Lambda$-dependencies in Fig.~\ref{fig:RvsL} for a generic operator. If we assume that $\mathcal{O}_{14_b}$ is responsible for neutrino masses, $\Lambda \sim 6\times 10^{4-5}$ TeV and we estimate that $\RME \simeq 10^{-38} - 10^{-36}$, further assuming $|g_{e\mu}|^2 = 1$. The current bound on $\RME < 1.7 \times 10^{-12}$ from the SINDRUM II collaboration, again assuming $|g_{e\mu}|^2 = 1$, implies $\Lambda \gtrsim 10$ GeV. As discussed in Sec.~\ref{sec:exp}, we expect the Mu2e experiment will be sensitive to $\RME \gtrsim 10^{-16}$ and is hence expected to observe $\mu^-\to e^+$ conversion if $\Lambda \lesssim 40$ GeV. 


\section{Results}
\label{sec:Results}

We follow the steps outlined for $\mathcal{O}_{14_b}$ in Sec.~\ref{sec:Estimates} and estimate the rates for $0\nu\beta\beta$ and $\mu^-\to e^+$ conversion for all effective operators listed in Tables~\ref{EstimateTableD5}, \ref{EstimateTableD7}, and \ref{EstimateTableD9}.  Analytic results are listed in Tables~\ref{EstimateTableD5}, \ref{EstimateTableD7}, and \ref{EstimateTableD9}.  The results for $0\nu\beta\beta$ agree with the estimates presented in Ref.~\cite{deGouvea:2007qla}, while the $\mu^-\to e^+$ conversion rates are the main results of this paper. 

In order to convert analytic expressions into numerical estimates for observables or the sensitivity to the new physics scale $\Lambda$,  we use the values listed in Table~\ref{table:Constants} for various SM parameters. $y_f$ denotes the Yukawa coupling of fermion $f$, $g$ denotes the weak gauge coupling, and $v$ denotes the vacuum expectation value of the Higgs field. The variables $Q$ -- different for $\RME$ and $\TBB$ -- were defined in Sec.~\ref{sec:Estimates} and are used to map our very rough estimates to more precise computations of $\RME$ and $\TBB$. We further assume that the operator coefficients $g_{\alpha\beta\dots}$ are all $\mathcal{O}(1)$. As discussed in Section~\ref{sec:theory}, this is not necessarily the case and should be kept under advisement. Several operators, e.g., $\mathcal{O}_{13}$, contain four or more leptons, and the resulting $0\nu\beta\beta$ and $\mu^-\to e^+$ amplitudes depend on a weighted sum of coefficients $g_{\alpha\beta\gamma\delta}$, typically of the form $\sum_\gamma g_{\alpha\beta\gamma\gamma} y_{l_\gamma}$, where $y_{l_\gamma}$ is the Yukawa coupling of the lepton of flavor $\gamma$. In these instances, we assume that $g_{\alpha\beta e e} \sim g_{\alpha\beta\mu\mu} \sim g_{\alpha\beta\tau\tau}$ and only list the largest contribution, usually due to the latter thanks to the relatively large tau Yukawa coupling. Numerical estimates for all observables under investigation are also listed in Tables~\ref{EstimateTableD5}, \ref{EstimateTableD7}, and \ref{EstimateTableD9}, assuming the operator in question is responsible for the observable neutrino masses, i.e., the value of $\Lambda$ agrees with the associated tabulated values of $\Lambda$. 
\begin{table}[!ht]
\caption{Constants used for estimating $\TBB$ and $\RME$.}
\begin{center}
\begin{tabular}{|c||c|c|c|c|c|c|c|c|c|c|}\hline
Constant & $G_F$ [GeV$^{-2}$] & $g$ & $\langle r\rangle$ & $v$ [GeV] & $Q$ [MeV] & $y_t$ & $y_b$ & $y_e$ & $y_\mu$ & $y_\tau$ \\ \hline \hline
Value & $1.17\times 10^{-5}$ & $0.653$ & $(100$ MeV$)^{-1}$ & $174$ & $11.0$ $(0\nu\beta\beta)$, $15.6$ $(\mu^-\to e^+)$ &  $0.9$ & $2\times 10^{-2}$ & $3\times 10^{-6}$ & $6\times 10^{-4}$ & $10^{-2}$ \\ \hline
\end{tabular}
\end{center}
\label{table:Constants}
\end{table}

\begin{figure}[!ht]
\centering
\includegraphics[width=\linewidth]{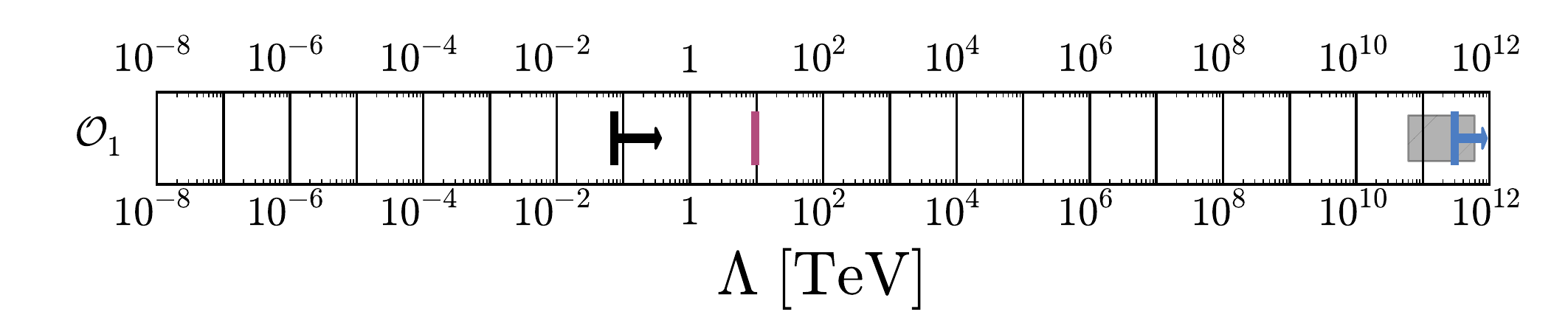}
\caption{Bounds on the effective scale associated with the dimension-five operator $\mathcal{O}_{1}$ from the KamLAND-Zen experiment for $0\nu\beta\beta$ (blue) and SINDRUM-II experiment for $\mu^-\to e^+$ (black). Also shown are the estimated sensitivity for the Mu2e experiment (pink) and the range of $\Lambda$ for which $m_{\alpha\beta} \sim 0.05 -0.5$ eV (grey). We assume $g_{\alpha\beta\ldots} = 1$ for all coefficients here. See text for details.}
\label{fig:ChartD5}
\end{figure}

\begin{figure}[!ht]
\centering
\includegraphics[width=\linewidth]{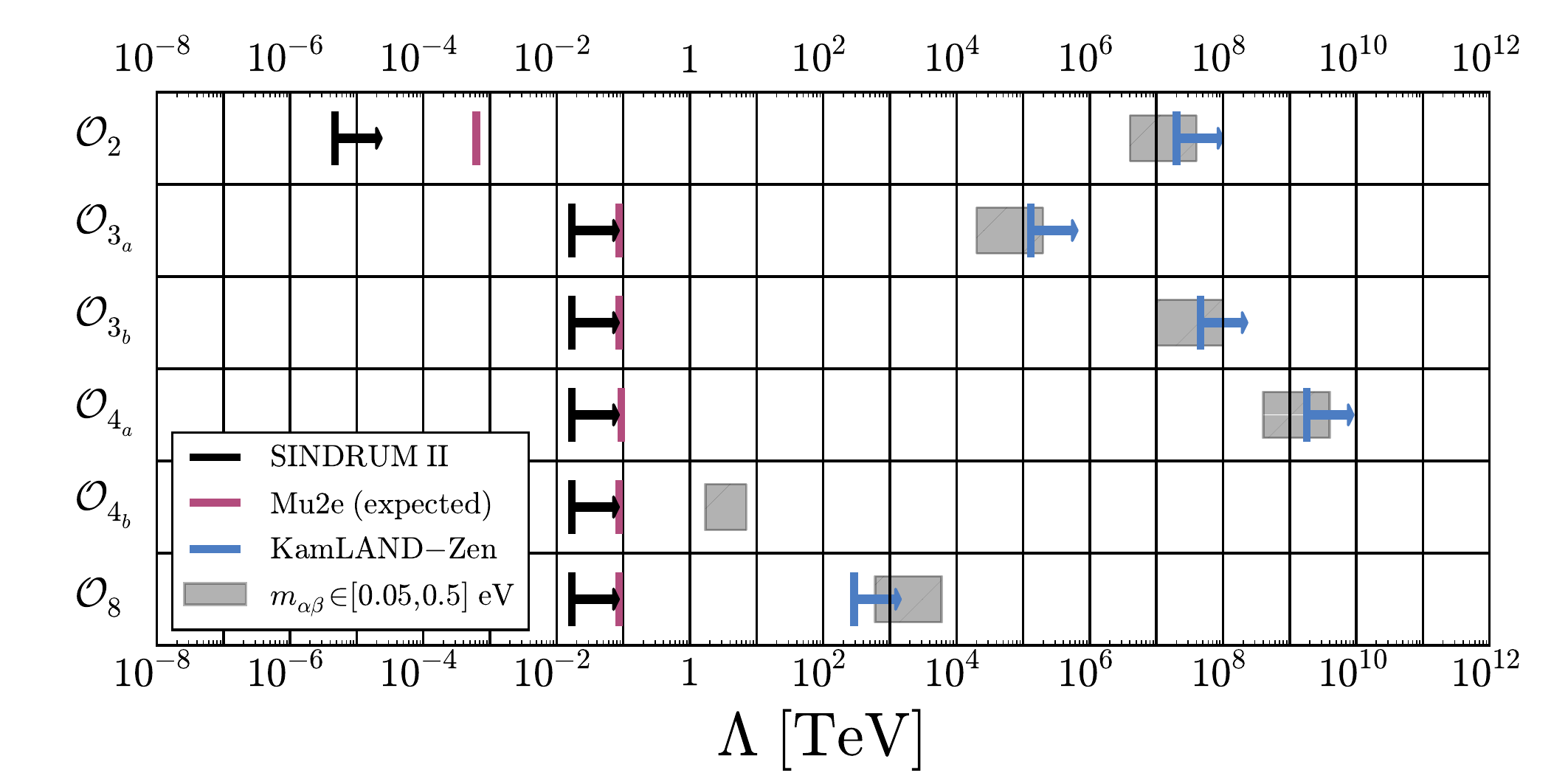}
\caption{Bounds on the effective scale associated with the dimension-seven operators $\mathcal{O}_{2,3_a,3_b,4_a,4_b,8}$ from the KamLAND-Zen experiment for $0\nu\beta\beta$ (blue) and SINDRUM-II experiment for $\mu^-\to e^+$ (black). Also shown are the estimated sensitivity of the Mu2e experiment (pink) and the range of $\Lambda$ for which $m_{\alpha\beta} \sim 0.05 -0.5$ eV (grey). We assume $g_{\alpha\beta\ldots} = 1$ for all coefficients here. See text for details.}
\label{fig:ChartD7}
\end{figure}

\begin{figure}[!ht]
\centering
\includegraphics[width=\linewidth]{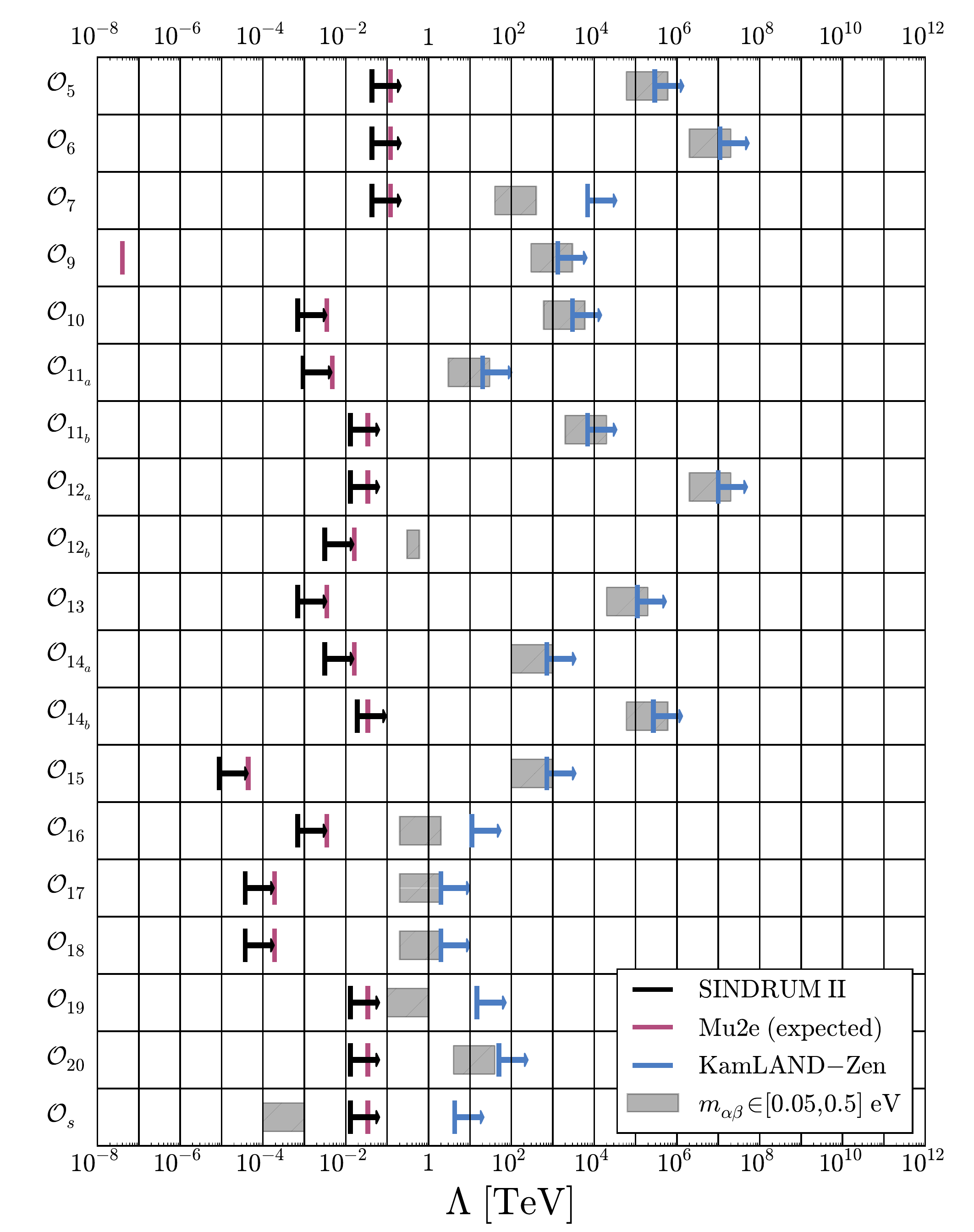}
\caption{Bounds on the effective scale associated with the dimension-nine operators listed in Table~\ref{EstimateTableD9} from the KamLAND-Zen experiment for $0\nu\beta\beta$ (blue) and SINDRUM-II experiment for $\mu^-\to e^+$ (black). Also shown are the estimated sensitivity of the Mu2e experiment (pink) and the range of $\Lambda$ for which $m_{\alpha\beta} \sim 0.05 -0.5$ eV (grey). We assume $g_{\alpha\beta\ldots} = 1$ for all coefficients here. See text for details.}
\label{fig:ChartD9}
\end{figure}

Figs.~\ref{fig:ChartD5},~\ref{fig:ChartD7}, and~\ref{fig:ChartD9} depict the currently allowed values of $\Lambda$ assuming current and future experimental bounds for operators of mass-dimension five, seven, and nine, respectively. For each operator, we depict the estimated bound for $\Lambda$ using the bound on $\RME$ from SINDRUM II (black), the estimated sensitivity of Mu2e (pink) and the estimated bound on $\Lambda$ using the results on $\TBB$ from the KamLAND-ZEN experiment (blue). All estimates are valid for  $g_{\alpha\beta\dots} = \mathcal{O}(1)$.  A hierarchical structure among the flavor coefficients could impair the ability to place a bound on $\Lambda$ from $0\nu\beta\beta$ or $\mu^-\to e^+$ conversion, for instance. Also shown for each operator is the range of $\Lambda$ such that $m_{\alpha\beta} \simeq 0.05-0.5$ eV.

As before, we direct the reader's attention to $\mathcal{O}_{4_b}$ and $\mathcal{O}_{12_b}$. These operators must have vanishing $g_{ee}$ due to the flavor-antisymmetry of the lepton doublets, meaning neither of these operators can produce $0\nu\beta\beta$, as indicated in Tables~\ref{EstimateTableD7} and \ref{EstimateTableD9}. There is no such restriction on $g_{e\mu}$. This means that, in principle, $\mu^- \to e^+$ conversion could occur at an observable rate in next-generation experiments in the complete absence of $0\nu\beta\beta$ if either of these operators were the only source of lepton-number violation. We note, however, that the neutrino mass matrices in Eqs.~\eqref{O4b-mass} and \eqref{O12b-mass} have vanishing diagonal elements, resulting in a mass matrix with relatively few independent degrees of freedom. This produces strong correlations among the neutrino masses and leptonic mixing parameters, such that current neutrino oscillation data preclude either of these operator from being the dominant contribution to neutrino masses and mixings (see, for instance, Refs.~\cite{Frampton:2002yf,Xing:2004ik,Singh:2016qcf}).

\section{Discussion and Conclusions}
\label{sec:Conclusions}
  
The observation of LNV phenomena would imply that the neutrinos are Majorana fermions and would help point the community to a subset of ideas for the new physics behind nonzero neutrino masses. The absence of LNV phenomena would not necessarily allow one to conclude that neutrinos are Dirac fermions, but a prolonged absence, assuming many different probes, would lead one to ultimately suspect this is the case and would point the search for the origin of nonzero neutrino masses down a very different path. Hence, deep and broad searches for the validity of lepton-number conservation are among the highest priorities of experimental particle physics today.

Here, we concentrated on understanding the reach of searches for $\mu^-\to e^+$ conversion in nuclei, partially motivated by the fact that, in the foreseeable future, several new experiments are aiming at improving the sensitivity to $\mu^-\to e^-$ conversion by four or more orders of magnitude. We opted for an effective operator approach that allows one to compare a large number of new physics scenarios. 

At face value, future searches for $\mu^-\to e^+$ conversion are sensitive to new, LNV physics at a wide range of effective energy scales, from 100~MeV to 10~TeV.\footnote{Except for ${\cal O}_9$. For $\Lambda\gtrsim 1$~GeV, the effective operator approaches is still valid for $\mu^-\to e^+$ conversion in nuclei, as long as the new physics is not very weakly coupled. It is, however, difficult to imagine that, for $\Lambda\lesssim100$~GeV, the existence of these new degrees of freedom is not severely constrained by probes of new phenomena that do not involve lepton-number violation. The exploration of such constraints cannot, however, be pursued within the formalism adopted here.}  This sensitivity pales in comparison with searches for $0\nu\beta\beta$, as revealed in Figs.~\ref{fig:ChartD5},~\ref{fig:ChartD7}, and~\ref{fig:ChartD9}. Comparisons between different LNV observables, however, need to be interpreted with care. Flavor effects can, as is well known, render the rate for $0\nu\beta\beta$ infinitesimally small and need not impact different LNV observables in the same way. 

LNV new physics ultimately leads to nonzero neutrino masses. Figs.~\ref{fig:ChartD5},~\ref{fig:ChartD7}, and~\ref{fig:ChartD9} also reveal that if the dominant contribution to the neutrino masses is captured individually by any of the effective operators discussed here, the expected rates for $\mu^-\to e^+$ conversion are well beyond the reach of next generation experiments, with one trivial exception.\footnote{If $\mathcal{O}_s$ was responsible for nonzero neutrino masses, its effective scale would be around 1~GeV and either $0\nu\beta\beta$-decay or $\mu^-\to e^+$ conversion should have been observed a long time ago, along with many more non-LNV observables.} All of these observations imply that, should $\mu^-\to e^+$ conversion be discovered in the next round of experiments, we will be able to conclude that (i) neutrinos are Majorana fermions, (ii) flavor effects, or something equivalent, significantly suppress the rate for $0\nu\beta\beta$, and (iii) the physics behind nonzero neutrino masses, assuming all new degrees of freedom are heavy, does not manifest itself at tree level via one of the effective operators investigated here but, instead, is captured by a non-trivial combination of operators whose contribution to the Majorana neutrino masses are significantly smaller than the contributions of any one operator. 

\vspace{5mm}

\textbf{Note added:} After this work was completed, Ref.~\cite{Geib:2016daa} appeared on the preprint archive. In it, the authors present a detailed calculation of the $0\nu\beta\beta$ decay rate for a model with a doubly-charged scalar, as well as a discussion of how to map specific models of new physics onto effective operator coefficients.


\begin{acknowledgments}
We thank Alex Merle for valuable feedback. AdG thanks Bob Bernstein for conversations and discussions that inspired some of this work. AK thanks Bill Molzon for useful conversations and feedback. The work of JMB, AdG, and KJK is supported in part by the DOE grant \#DE-SC0010143. The work of AK is supported in part by DOE grant \#DE-SC0009919.
\end{acknowledgments}

\bibliographystyle{apsrev-title}
\bibliography{mue}{}

\end{document}